\documentclass[final,5p,times,twocolumn,authoryear]{elsarticle}

\usepackage{natbib}
\usepackage{amsmath,amssymb,amsfonts}
\usepackage{colortbl}
\usepackage{pifont}
\newcommand{\cmark}{\ding{51}}
\newcommand{\xmark}{\ding{55}}
\usepackage{algorithmic}
\usepackage{graphicx}
\usepackage{tikz}
\usepackage{float}
\usepackage{multirow}
\usepackage{textcomp}
\usepackage{xspace}
\usepackage{bm}
\usepackage{hyperref}
\usepackage{scalerel}
\usepackage{booktabs}
\usepackage{longtable}
\usepackage{array}
\usepackage{threeparttable}
\usepackage{rotating}
\usepackage{lscape}
\usepackage{tabularx}
\usepackage[table]{xcolor}
\definecolor{LightGray}{gray}{0.95}
\definecolor{purple}{HTML}{9CA3D6}
\definecolor{pink}{HTML}{edcdea}
\definecolor{red}{HTML}{FF8B94}
\definecolor{green}{HTML}{FFD3B6}
\definecolor{orange}{HTML}{c7ddba}
\definecolor{blue}{HTML}{648fff}
\definecolor{robcol}{RGB}{227,153,216}
\definecolor{statcol}{RGB}{186,237,184}
\definecolor{unccol}{RGB}{140,206,232}
\definecolor{repcol}{HTML}{F5CA97}
\definecolor{bggray}{RGB}{230,230,230}
\newcommand{\yes}{\raisebox{-.3ex}{\textbullet}}
\newcommand{\minibar}[5][]{
  \def\lbfont{\scriptsize}
  \ifx b#1\def\lbfont{\scriptsize\bfseries}\fi
  \pgfmathsetmacro{\barw}{0.87\linewidth}
  \begin{tikzpicture}[baseline=0.6ex]
    \fill[bggray] (0,0) rectangle (\barw pt, 5pt);
    \fill[#2]     (0,0) rectangle ({#5/100*\barw pt}, 5pt);
    \node[anchor=west, font=\lbfont, text=black]
      at ({#5/100*\barw pt + 2pt}, 2.5pt) {#3 / #4 (#5\%)};
  \end{tikzpicture}%
}

\journal{Medical Image Analysis}

\begin{document}

\begin{frontmatter}



\title{Data-Driven Registration and Modeling of Brain Deformation for Image-Guided Neurosurgery}


\author[first]{Tiago Assis\fnref{label1}}
\ead{tassis@lasige.di.fc.ul.pt}
\author[second]{Colin P.~Galvin}
\author[third]{Joshua P.~Castillo}
\author[second]{Nazim Haouchine}
\author[third]{Marta Kersten-Oertel}
\author[fourth,fifth]{Zeyu Gao}
\author[fourth,fifth]{Mireia Crispin-Ortuzar}
\author[sixth]{Stephen J.~Price}
\author[sixth]{Thomas Santarius}
\author[seventh]{Yangming Ou}
\author[second]{Sarah Frisken}
\author[first]{Nuno C.~Garcia}
\author[second]{Alexandra J.~Golby}
\author[eighth]{Reuben Dorent}
\author[fourth,fifth]{Ines P.~Machado\corref{cor1}}
\ead{im549@cam.ac.uk}

\fntext[label1]{Present address: Institute for Systems and Robotics, Laboratory for Robotics and Engineering Systems (LARSyS), Instituto Superior Técnico, University of Lisbon, 1049-001 Lisbon, Portugal.}
\cortext[cor1]{Corresponding author.}

\affiliation[first]{organization={LASIGE, Faculty of Sciences, University of Lisbon},
            postcode={1749-016 Lisbon},
            country={Portugal}}
\affiliation[second]{organization={Department of Neurosurgery and Department of Radiology, Brigham and Women's Hospital, Harvard Medical School},
            city={Boston},
            postcode={MA 02115},
            country={USA}}
\affiliation[third]{organization={Gina Cody School of Engineering and Computer Science, Concordia University},
            city={Montreal},
            postcode={QC H3G 2W1},
            country={Canada}}
\affiliation[fourth]{organization={Cancer Research UK Cambridge Centre, University of Cambridge},
            city={Cambridge},
            postcode={CB2 0RE},
            country={UK}}
\affiliation[fifth]{organization={Department of Oncology, University of Cambridge},
            city={Cambridge},
            postcode={CB2 0AH},
            country={UK}}
\affiliation[sixth]{organization={Department of Clinical Neurosciences, University of Cambridge},
            city={Cambridge},
            postcode={CB2 0QQ},
            country={UK}}
\affiliation[seventh]{organization={Department of Radiology, Boston Children's Hospital, Harvard Medical School},
            city={Boston},
            postcode={MA 02115},
            country={USA}}
\affiliation[eighth]{organization={Sorbonne Université, Institut du Cerveau - Paris Brain Institute - ICM},
            addressline={CNRS, Inria, Inserm, AP-HP, Hôpital de la Pitié Salpêtrière},
            postcode={75013 Paris},
            country={France}}

\begin{abstract}
Accurate compensation of brain deformation is critical for reliable image-guided neurosurgery. Surgical manipulation and tumor resection induce tissue motion, causing preoperative planning images to become misaligned with the intraoperative anatomy. In this review, we examine data-driven methods developed between 2020 and 2025 for brain deformation registration and modeling, with a particular focus on learning-based approaches. A comprehensive literature search was conducted in PubMed, IEEE Xplore, Scopus, and Web of Science using predefined inclusion and exclusion criteria for computational methods addressing brain deformation in neurosurgical imaging, resulting in 46 eligible studies. We provide a unified analysis of methodological strategies, including deep learning-based image registration, direct deformation field regression, synthesis-driven multimodal alignment, resection-aware architectures for handling missing correspondences, and hybrid models integrating biomechanical priors. We also examine dataset utilization, evaluation metrics, validation protocols, and the assessment of uncertainty and generalization across studies. While learning-based methods demonstrate promising accuracy and computational efficiency, current approaches remain limited by out-of-distribution robustness, standardized benchmarking, interpretability, and readiness for clinical deployment. Our review highlights these gaps and outlines future directions toward more robust, generalizable, and clinically translatable solutions for neurosurgical guidance. By organizing recent advances and critically assessing evaluation practices, this work provides a comprehensive reference for researchers and clinicians working on data-driven registration and modeling of brain deformation.
\end{abstract}




\end{frontmatter}




\section{Introduction}
\label{sec:introduction}
The brain is responsible for controlling motor, sensory, visual, language, and other cognitive and emotional functions that underlie human behavior. For this reason, surgical procedures involving the brain are highly complex and carry significant risks, since brain tumors are often located near critical structures. One of the greatest challenges in this type of surgery is distinguishing between the tumor and healthy brain tissue. To the naked eye, they can appear almost identical, meaning surgeons may inadvertently touch or remove areas responsible for essential functions such as language or movement, mistaking them for tumor tissue. Consequently, the primary goal of neurosurgery is to maximize tumor removal while minimizing damage to surrounding tissue. 

In recent decades, image-guided systems have revolutionized neurosurgical practice, providing surgeons with real-time anatomical guidance to assist in the maximal safe resection of surgical
targets \citep{asfaw2024charting}. These systems rely on patient-to-image registration algorithms to map the physical patient in the operating room (OR) to the patient's preoperative images, such
as computed tomography (CT), magnetic resonance imaging (MRI), functional MRI (fMRI), or diffusion tensor imaging (DTI), enabling intraoperative visualization and planning of the incision,
craniotomy, and resection.

Despite their utility, these systems are limited by a phenomenon known as \textit{brain shift}, which invalidates the rigid body assumption inherent to most image-guided navigation systems. Brain
shift is defined as brain deformation that results from physical, surgical, or biological factors that occur during surgery. It results in discrepancies between the preoperative imaging and
the actual brain anatomy, which tend to increase as surgery progresses. The causes of brain shift are multifactorial, influenced by lesion location and pathology, edema, cerebrospinal fluid (CSF)
loss, gravity, and tissue resection \citep{dorward1999postimaging,ohue2010evaluation,gerard2017brain,fan2022comparison}. Drugs such as mannitol and dexamethasone, used to reduce swelling and
promote brain relaxation, can also lead to brain deformation \citep{peng2014effect,de2024hypertonic}. Without accurate, real-time guidance during surgery, brain shift and the difficulty of
distinguishing tumor tissue from healthy tissue can result in incomplete tumor removal or accidental damage to critical brain structures, leading to higher rates of tumor recurrence, neurological
deficits such as speech, motor, or memory impairments, longer recovery times, and ultimately reduced survival rates and quality of life for patients
\citep{lacroix2001multivariate,sommer2015resection,lara2017advances}.

Recent advances in deep learning for medical image registration, along with the increasing availability of large public datasets \citep{baheti2024brain,juvekar2024remind}, show promise for modeling tissue deformation during surgery by identifying patterns of brain shift and suggesting optimal compensation strategies. These advances span image registration \citep{dorent2025brain,hansen2025learn2reg}, deformation modeling \citep{el2024hyperu,assis2025deep}, and the prediction of intraoperative and postoperative imaging in neurosurgery \citep{haouchine2023learning}.

The growing body of literature from these datasets and challenges suggests that deep learning-based methods for brain deformation compensation may become a key component of neurosurgical image
guidance. Two key reviews, \citet{gerard2017brain} and \citet{gerard2021brain}, summarized earlier methods for brain shift correction. The first provides a broad overview of the causes,
quantification, and compensation strategies for brain shift, using a taxonomy of physical, surgical, and biological factors. The second focuses on developments in intraoperative ultrasound
(iUS)-based approaches from 2015 to 2020, highlighting trends in mathematical modeling and the growing interest in vessel-based registration, as well as integration with augmented and virtual
reality technologies. Complementarily, the present review adopts a neurosurgery-specific, method-centric perspective on recent learning-based approaches for brain deformation
compensation. We focus on the current landscape of data-driven deformation modeling techniques developed since 2020, reviewing key algorithmic strategies including deep learning-based
registration, synthesis-driven multimodal alignment, resection-aware architectures, and hybrid physics-informed frameworks. Beyond algorithmic categorization, we critically examine how these
methods are evaluated, including practices for uncertainty quantification, validation and benchmarking, robustness to resection-induced topology changes, and constraints imposed by intraoperative
deployment. Our goal is to provide a coherent analytical foundation that links methodological design choices to clinical applicability, while highlighting open challenges and opportunities for
future research and clinical deployment.

\section{Background}
In neurosurgery, accurate real-time visualization of brain anatomy is enabled by intraoperative imaging and supported by computational approaches that account for intraoperative brain deformation. This section provides an overview of the neurosurgical workflow (Section~\ref{sec:neuro_workflow}) and the main intraoperative imaging approaches, highlighting their strengths and limitations (Section~\ref{sec:intraop_imaging}). It also introduces complementary strategies for brain shift compensation, including image-based registration (Section~\ref{sec:image_registration_background}), biomechanical modeling (Section~\ref{sec:biomech_modeling_background}), and physics-guided neural networks (Section~\ref{sec:pinns_background}).

\subsection{Neurosurgical workflow}
\label{sec:neuro_workflow}

\begin{figure*}[!t]
    \centering
    \includegraphics[width=\textwidth,height=\textheight,keepaspectratio]{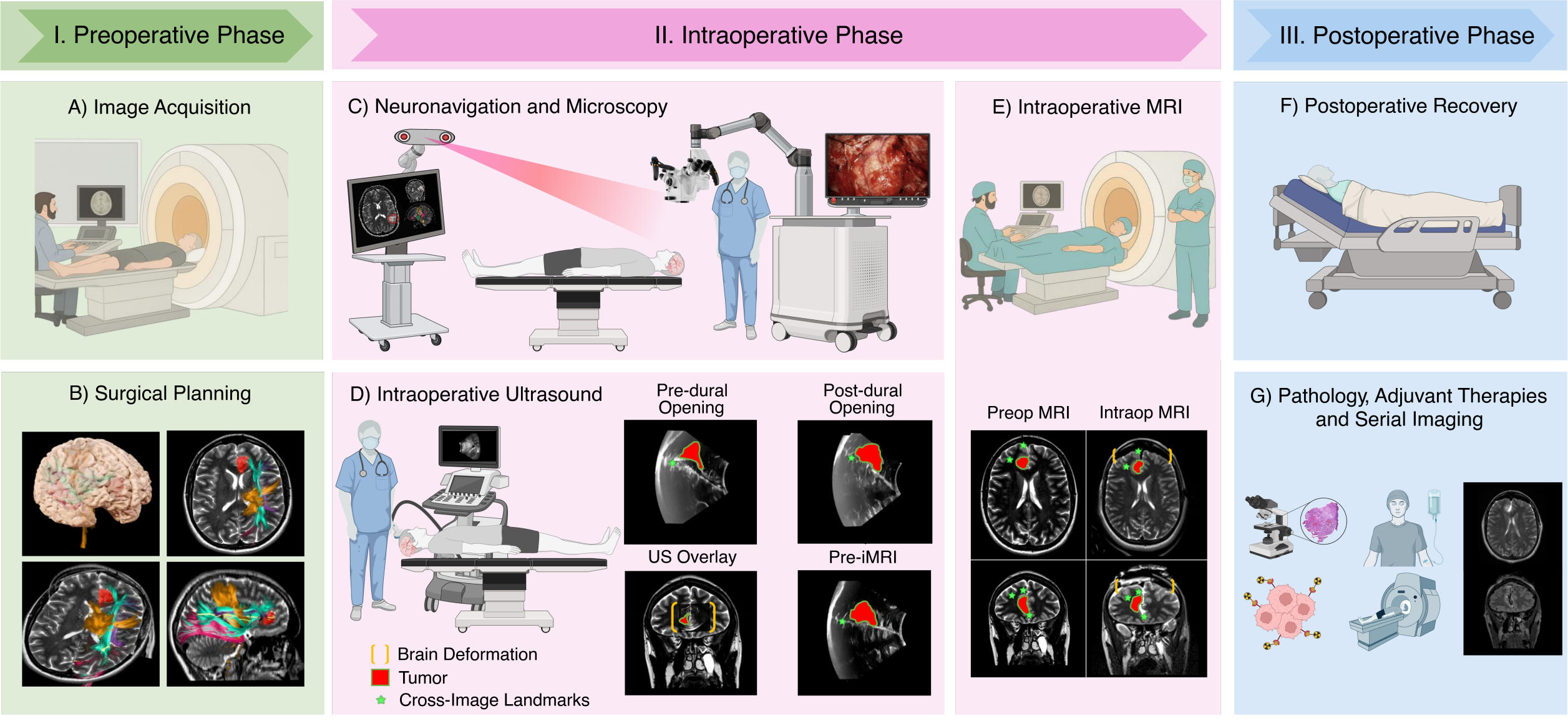}
    \caption{Neurosurgical workflow including preoperative assessment, intraoperative phase, and postoperative care (Phases I, II, and III, respectively). (A) \textit{Image Acquisition}: Following clinical evaluation and discussion of treatment options, preoperative imaging is acquired. (B) \textit{Surgical Planning}: Multiparametric magnetic resonance imaging (MRI) is the primary imaging modality used to delineate tumor boundaries and assess the proximity of eloquent and critical brain areas. Additional modalities, such as CT, functional MRI, or diffusion MRI, may be incorporated when needed to guide individualized surgical plans and resection goals. (C) \textit{Neuronavigation and Microscopy}: In the operating room, patient-to-image registration aligns the patient's anatomy with the preoperative imaging, establishing a common coordinate system for neuronavigation. The incision and craniotomy are planned and executed to provide surgical access to the lesion, while neuronavigation and intraoperative microscopy provide continuous image-guided visualization throughout the resection. (D) \textit{Intraoperative Ultrasound}: Ultrasound is employed at critical points: pre-dural to confirm tumor access, post-dural to delineate margins and assess brain shift, and at intervals during resection to monitor for residual tumor. (E) \textit{Intraoperative MRI}: Intraoperative MRI assesses the extent of resection and brain shift. If residual tumor is present, additional targeted resection is guided by updated MRI data alongside ultrasound and microscopy, before closure. Green stars indicate the same location across imaging modalities and time points. (F) \textit{Postoperative Recovery}: Patients undergo immediate postoperative clinical care and imaging to evaluate the final extent of resection and screen for postoperative complications. (G) \textit{Pathology, Adjuvant Therapies and Serial Imaging}: Pathology informs adjuvant therapy decisions, and serial MRI with ongoing clinical follow-up monitors for recurrence and optimizes long-term management.}
    \label{fig:fig1}
\end{figure*}

A schematic overview of the neurosurgical workflow is shown in Fig.~\ref{fig:fig1}, illustrating the progression from preoperative assessment and multimodal imaging (Phase I) to intraoperative
navigation, visualization, and tumor resection (Phase II), followed by postoperative imaging and follow-up care (Phase III). Before entering the OR, careful preoperative planning brings together
multiple imaging methods, such as MRI, CT, fMRI, and DTI, to build a detailed map of the patient's brain (Phase I.B) \citep{menze2014multimodal,juvekar2023mapping}. Once the patient is positioned
and the head is secured in a rigid clamp, registration is performed to align the patient's anatomy with preoperative imaging, allowing real-time tracking of instruments during the procedure.
Registration is typically based on identifying anatomical landmarks or surface points and may be refined using fiducial markers or surface-matching techniques. At the start of a case, a typical
target registration error (TRE) of $1$--$2$~mm is achievable; however, brain shift can cause this to drift to $4$--$6$~mm as surgery progresses, with displacements of the
cortical surface exceeding $20$~mm reported in long resections \citep{stieglitz2013silent,gerard2021brain}. Most neuronavigation platforms target a TRE of $2$--$3$~mm, and registration is
considered acceptable when the measured registration error falls within a predefined range and the surgeon confirms alignment before surgery proceeds \citep{taleb2023image}. Once registration is
confirmed, the surgeon plans the incision and craniotomy to reach the target safely while limiting tissue disruption.

Intraoperative visualization relies on a combination of optical and imaging technologies that guide accurate and safe resection. The operating microscope provides high magnification and
illumination for detailed visualization and careful dissection (Phase II.C). In parallel, iUS offers real-time imaging that complements static preoperative scans and can be co-registered with
navigation images to localize the target, monitor brain shift, and assess the extent of resection at key surgical checkpoints (Phase II.D) \citep{sass2022navigated}. To maintain navigation
accuracy, updates are ideally integrated at critical points: post-dural opening, mid-resection, and pre-closure or intraoperative MRI (iMRI), with a practical target of $30$--$60$ seconds per update
while maintaining sub-$3$~mm accuracy  \citep{haouchine2023learning,rahmani2024d2bgan}. Intraoperative MRI can further assess the extent of resection and identify residual tissue, with
updated images co-registered into the navigation system to compensate for brain shift and maintain accurate guidance (Phase II.E) \citep{ginat20143,noh2020intraoperative}. However, given the cost
and logistical limitations of iMRI, it is typically reserved for a single acquisition near the end of the planned resection rather than used for continuous navigation updates. Early postoperative
imaging documents the extent of resection, screens for complications, and, together with histopathologic and molecular findings, guides further therapy and long-term follow-up (Phase III.G).

\subsection{Intraoperative imaging and visualization modalities}
\label{sec:intraop_imaging}
Intraoperative MRI is considered the gold standard for updating registration during surgery, enabling visualization of residual tumor and critical structures while effectively restoring
navigational accuracy to the current intraoperative anatomy \citep{ginat20143,noh2020intraoperative,zaffino2020review,matsumae2022intraoperative}. The newly co-registered imaging acquired in
familiar contrasts such as T1, contrast-enhanced T1 (ceT1), T2, and FLAIR, provide surgeons with an intuitive and detailed update of the surgical target in standard orthogonal planes. Similarly,
intraoperative CT (iCT) can be used to update registration and inform stereotactic procedures, but is constrained by radiation exposure and limited soft tissue contrast, restricting its routine
application \citep{ashraf2025value}. Although the adoption of iMRI has been increasing every year for the past decade, it is still limited by high costs and substantial logistical demands, as iMRI
installations can exceed USD $5$ million each once shielding and instrumentation are included, thereby confining adoption to well-funded centers \citep{factmr2026imri}. Intraoperative US
provides a widely available and cost-effective option for real-time surgeon-guided intraoperative imaging \citep{dixon2022intraoperative}. When paired with commercial navigation systems, it can be
registered to the preoperative or intraoperative MRI, allowing assessment of tumor resection as it progresses and informing surgeons of brain shift in real time
\citep{bastos2021challenges,mosteiro2022intraoperative,sass2022navigated,wei2023application}. However, iUS also has notable limitations: image quality can be variable, interpretation requires
experience, and outcomes are influenced by interoperator variability \citep{bastos2021challenges,dixon2022intraoperative}, greatly influencing its utility and reliability when updating
registration.

In addition to dedicated intraoperative imaging, visualization technologies such as high-resolution surgical microscopes and Augmented Reality (AR) overlays provide intraoperative feedback by
integrating imaging data directly into the surgical field \citep{hey2023augmented}. However, these systems do not inherently compensate for brain shift and currently play a limited role in
updating registration. Analogous to how iUS can be registered to preoperative MRI to provide real-time deformation updates, emerging work has begun to use the surgical microscope as a registration
tool from cortical surface features and vessel landmarks to estimate and compensate for brain shift directly from the operative field
\citep{haouchine2023learning,haouchine2021pose,haouchine2021predicted}.

\subsection{Brain deformation compensation approaches}
Brain deformation compensation has traditionally been addressed using two main strategies: image-based registration methods and physics-based biomechanical modeling \citep{sotiras2013deformable}.
Both aim to update preoperative plans to the intraoperative brain anatomy, but they differ in how they estimate deformations. This review focuses primarily on image-based registration,
particularly recent advances in machine learning and deep learning (DL), while also highlighting biomechanics-based methods for their complementary strengths and their growing integration with
data-driven frameworks. Fig.~\ref{fig:classic_vs_dl} compares classic instance optimization (IO)-based registration with DL-based frameworks, highlighting their differences in how deformation fields
are estimated and optimized.

\subsubsection{Image-based registration}
\label{sec:image_registration_background}
\begin{figure*}[!t]   
    \centering    
    \includegraphics[width=\textwidth,keepaspectratio]{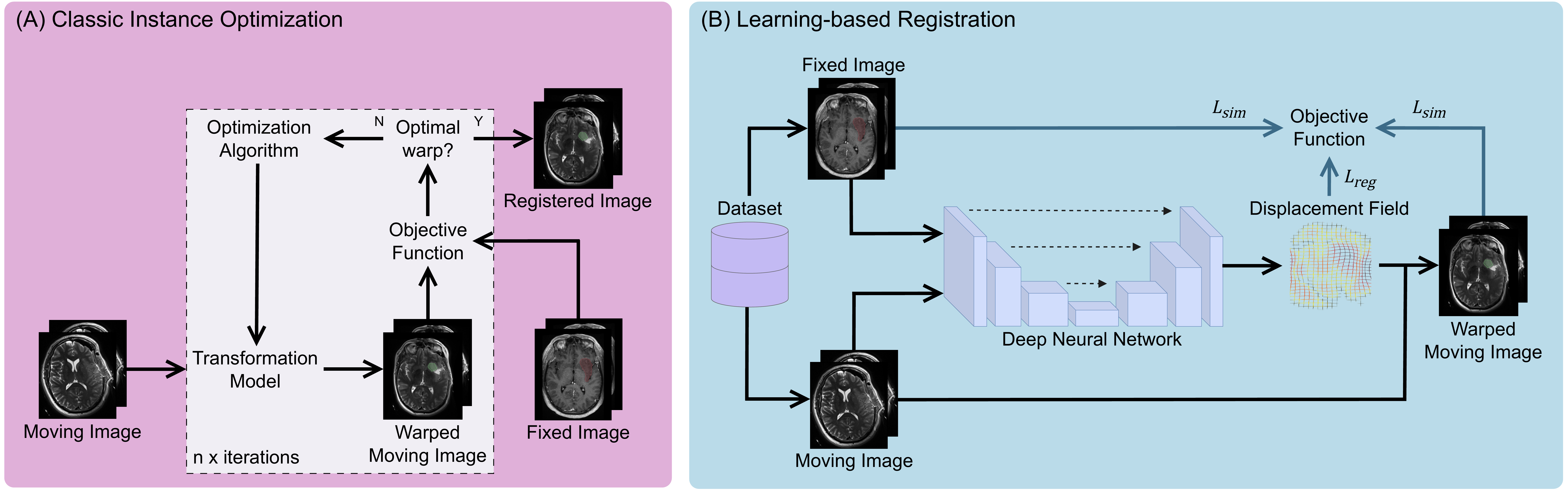}
    \caption{Comparison between classic instance optimization and deep learning-based medical image registration frameworks. (A) Classic methods optimize the transformation parameters independently for each image pair by minimizing a similarity-based objective function through iterative optimization (Section~\ref{sec:classic_instance_optimization}). (B) Learning-based approaches train a neural network on a dataset of image pairs to predict the deformation field directly in a single forward pass, minimizing both similarity ($L_{sim}$) and regularization ($L_{reg}$) losses during training (Section~\ref{sec:learning_based_registration}). Blue arrows show the inputs to the objective function that guides backpropagation. Tumor segmentations are shown in green and red on the moving and fixed images, respectively.}
    \label{fig:classic_vs_dl}
\end{figure*}

The primary objective of image registration is to accurately capture the deformation between two images, formulated as an optimization problem. Conventional image registration methods
(Fig.~\ref{fig:classic_vs_dl}-A) estimate spatial transformations between image pairs by iteratively optimizing a predefined objective function that relies on handcrafted similarity metrics
\citep{maes2002multimodality,sarvaiya2009image,hjelm2018learning,liu20222d} and regularization terms to enforce smooth and physically plausible deformations \citep{reithmeir2024data} (see Sections~\ref{sec:similarity_measures} and \ref{sec:regularization_terms}). Parameter tuning is typically manual, requiring domain expertise to balance accuracy and robustness. Since the optimization is
performed independently for each image pair, these methods tend to be computationally expensive and sensitive to initialization. Deep learning-based methods (Fig.~\ref{fig:classic_vs_dl}-B) shift this
paradigm by learning the mapping from image pairs to displacement fields across datasets, allowing the optimization of the transformation to be guided by training data rather than manual
configuration for each case \citep{chen2025survey} (Section~\ref{sec:image_registration}). Once trained, these models can register unseen image pairs in real time. However, DL-based methods have their
own limitations; for instance, they require large annotated or paired datasets, which are scarce in neurosurgical contexts, and tend to generalize poorly to data drawn from outside their training
distribution. Performance can thus degrade in real intraoperative conditions involving severe deformation, resection cavities, or multimodal inputs not represented during training. These
limitations motivate the hybrid and biomechanics-guided strategies discussed in Sections~\ref{sec:hybrid_learning} and \ref{sec:biomech_modeling_physics_registration}.

\subsubsection{Biomechanical modeling}
\label{sec:biomech_modeling_background}
Modeling brain deformation with biomechanics involves formulating a physical model defined by brain geometry, tissue material properties, loading forces, and boundary conditions (BCs), while
incorporating physics loads, such as changes in intracranial pressure due to dural opening, CSF drainage, and the influence of gravity
\citep{mostayed2013biomechanical,frisken2019intra,lesage2021viscoelastic,yu2022automatic}. The geometric model typically includes the brain, skull, ventricles, and tumors segmented from
preoperative imaging. Material properties are usually derived from population-based studies or elastography \citep{miller2000mechanical,bilston2019brain}. Loading and BCs are generally specified
at the brain-skull interface, along with mechanically anchored regions (e.g.,~the brainstem), as well as external forces such as surgical instruments, CSF drainage, and gravity
\citep{mazumder2013mechanical,miller2019biomechanical}. This physical system is governed by partial differential equations (PDEs) derived from continuum mechanics and incorporating the chosen
constitutive laws. These PDEs are typically solved using the finite element method (FEM), which offers a flexible and robust numerical framework for handling complex geometries and heterogeneous
material properties in brain biomechanics, although meshless discretization approaches have also been investigated \citep{zhang2013patient,yu2022automatic}.

\subsubsection{Physics-guided deep learning}
\label{sec:pinns_background}
Although the two have traditionally been treated independently, recent efforts have begun to integrate image registration with biomechanical modeling. Physics-informed neural networks (PINNs) or DL models that
learn from biomechanical priors aim to retain the speed of data-driven methods while ensuring anatomical and physical plausibility. Common strategies include using biomechanical models to generate
synthetic ground truth deformations for supervised DL-based registration approaches \citep{assis2025deep}, incorporating PDE-based loss terms \citep{min2024biomechanics}, using FEM outputs as
priors or regularizers \citep{hu2018adversarial}, or directly driving FEM-based models with sparse intraoperative information such as matched keypoints \citep{frisken2020comparison}. These hybrid
methods represent promising avenues for bridging the performance and generalizability gap in brain deformation compensation. Section~\ref{sec:biomech_modeling_physics_registration} examines methods for
biomechanical modeling of the brain and physics-guided intraoperative registration.

\subsection{Algorithmic components in image registration} 
At the heart of image registration lies the search for a transformation that best aligns two images within a shared spatial domain. The following sections introduce the mathematical formulation of this problem (Section~\ref{sec:formulation}) and the key components that govern registration, including similarity measures (Section~\ref{sec:similarity_measures}), which quantify correspondence between moving (source) and fixed (target) images, and regularization terms (Section~\ref{sec:regularization_terms}), which enforce physically plausible deformations.

\subsubsection{Problem formulation}
\label{sec:formulation}
In a 3D image space $\Omega \subset \mathbb{R}^3$, we assume a moving image $I_M \in \mathbb{R}^{D \times H\times W}$ that requires alignment with a fixed image $I_F \in \mathbb{R}^{D' \times H'\times W'}$, where $D$ is the depth, $H$ is the height, and $W$ is the width. The goal of
registration is to estimate the optimal transformation $\phi : \Omega_F \rightarrow \Omega_M$ that will register both images using a composition $\circ$ operation, so that the dissimilarity between $I_F$ and
$I_M \circ \, \phi$ is minimized. The deformation space depends on the approach, and the estimated transformation is usually either affine or deformable. If affine, a transformation matrix displaces the
image globally, according to a combination of rotation, translation, scaling, and shearing parameters. When the transformation is deformable, it can be represented either: (1) as a dense
displacement vector field, which individually maps each voxel to a new spatial location; or (2) in a compact parametric form, such as spline-based models, where the deformation is defined by a
sparse set of control points and smoothly interpolated over the image domain \citep{rueckert2010medical,sotiras2013deformable}. Deep learning methods can employ any of these formulations to
predict the optimal affine parameters or dense displacement field $\hat{\phi}$, using an objective function that is optimized over the possible domain of transformations. Formally, it can be expressed as:
\begin{equation} 
    \label{eq:obj_function}
    \hat{\phi} = \operatorname*{arg\min}_\phi \: - \mathcal{S}(I_{F}, I_{M}\circ\,\phi) + \lambda \, \mathcal{R}(\phi),
\end{equation}
where $\mathcal{S}$ and $\mathcal{R}$ are the similarity and regularization terms, respectively, and $\lambda$ weights the regularization.

\subsubsection{Similarity measures}
\label{sec:similarity_measures}
The selection of an appropriate similarity measure depends on the image modalities involved, the nature of the registration, and the application requirements. In general, similarity measures can be classified into intensity-based and feature-based measures, each offering distinct advantages and limitations. 

\textbf{Intensity-based measures.} Intensity-based measures compare voxel values directly and work best when corresponding structures exhibit consistent intensity patterns across images. These
measures typically capture intensity differences, correlations, or probabilistic dependencies \citep{rueckert2010medical,oliveira2014medical}. For instance, Mean Squared Error (MSE) quantifies the
average voxel-level intensity differences between images.
MSE assumes that corresponding structures in the image have identical intensity distributions and is sensitive to intensity variations caused by image artifacts, as well as cross-platform and
cross-modality image intensity variations. Thus, MSE has limited applicability in multimodal imaging. Normalized Cross-Correlation (NCC) assumes a linear relationship between image intensities
while eliminating the effects of variations in brightness and contrast (amplitude of the signal) \citep{avants2008symmetric}. 
Local NCC (LNCC) is an extension that computes the NCC within localized neighboring regions of each voxel, making it more robust to intensity variations and nonlinear intensity mappings across the
entire image domain. This is particularly useful in deformable image registration, where complex anatomical changes are common. Mutual Information (MI) is an intensity-based measure that can be
effective in multimodal registration \citep{viola1997alignment}. Methods that use MI are based on the joint intensity distribution of the two images, which captures the statistical relationship
between their voxel intensities. MI quantifies the reduction in joint entropy achieved through alignment. When images are correctly aligned, their joint intensity distributions exhibit stronger
statistical dependence, leading to a maximization of MI.
As in the case of LNCC, measures based on information theory can also be locally computed to improve robustness by accounting for localized regions.
Nevertheless, MI tends to underperform in the presence of large modality gaps, since various intensity configurations can produce similarly low joint entropy \citep{wein2008automatic}. To address
this limitation, the Linear Correlation of Linear Combination ($\text{LC}^{2}$) measure \citep{wein2008automatic} was introduced to better handle substantial domain shifts. Defined at the patch level
and inspired by US image formation physics, $\text{LC}^{2}$ has demonstrated competitive performance for US-to-MR and US-to-CT registration tasks \citep{wein2008automatic,fuerst2014automatic}.
Although $\text{LC}^{2}$ is inherently non-differentiable, \citet{ronchetti2023disa} proposed a DL approach to approximate this measure, enabling its integration into end-to-end trainable frameworks. Another
widely adopted gradient-based measure is Normalized Gradient Fields (NGF) \citep{haber2006intensity}, which compares the orientation of intensity gradients between images rather than intensity
values. By being invariant to intensity scale and to the sign of contrast inversions, NGF is well suited to multimodal alignment.

\textbf{Feature-based measures.} Feature-based registration methods rely on detecting corresponding anatomical features, such as keypoints or local descriptors, in both the fixed and moving images
\citep{luo2018using,luo2018feature,machado2018non,machado2019deformable,joutard2022driving,wang2023robust}. The transformation is then estimated by minimizing the spatial discrepancy between
these matched features. Unlike intensity-based measures, which operate on voxel intensity values, feature-based methods use higher-level representations that capture structural or contextual cues.
This makes feature-based registration robust to large deformations, partial fields of view, and topological changes, as well as being especially beneficial in multimodal registration, where
intensity relationships may be inconsistent or unreliable \citep{sotiras2013deformable}.

Traditional keypoint detector and descriptor methods \citep{lowe2004distinctive,bay2006surf,leutenegger2011brisk} have been adapted to 3D medical images, while others have been specifically handcrafted for medical image registration, such as the Modality Independent Neighborhood Descriptor (MIND) \citep{heinrich2012mind}, which identifies
regions based on self-similarity \citep{heinrich2013towards} within a local neighborhood around a keypoint. Deep learning approaches have also shown strong performance in detecting keypoints and
computing descriptors tailored to specific anatomical contexts, commonly leveraging convolutional neural networks (CNNs) \citep{detone2018superpoint,yi2016lift}. Several novel works have also
continued to focus on designing descriptors and networks that are contrast-invariant and anatomically meaningful for multimodal neurosurgical applications. \citet{rasheed2024learning} proposed
learning contrast-invariant descriptors for MRI-iUS registration using a synthesis-based contrastive learning framework, and \citet{morozov20253d} extended this concept to a 3D setting by
learning descriptors for MR-iUS volumes that enable patch-based rigid registration using a sparse set of keypoint matches. Similarly, \citet{salari2023towards} further demonstrated how contrastive
learning can produce MRI-iUS keypoint detectors that align with clinically meaningful structures such as vessels and sulci.

\textbf{Additional anatomical information.} Anatomical priors, such as segmentations and landmarks, serve as valuable sources of structural guidance that complement intensity-based information during image registration, constraining and informing the alignment process. Segmentation masks, when available, enable registration methods to focus on aligning meaningful anatomical contours rather than relying solely on voxel intensities. A common way to incorporate segmentations is by measuring the overlap between the warped segmentation from the moving image and the fixed image segmentation using metrics such as the Dice score and Hausdorff distance (HD). Complementarily, landmarks can directly guide an intensity-based registration process by serving as a ground truth signal. The TRE is the standard measure for assessing how precisely anatomical points are aligned, defined as the Euclidean distance between the transformed moving landmark and its fixed image counterpart.

\subsubsection{Regularization terms}
\label{sec:regularization_terms}
Image registration is an inherently ill-posed problem since multiple transformations may map one image to another with comparable accuracy while resulting in anatomically or physically implausible
deformations. To address this problem, regularization of the optimization problem (Eq.~\ref{eq:obj_function}) is necessary to introduce prior knowledge and constraints that model desirable
deformation properties (e.g.,~smoothness and invertibility) \citep{reithmeir2024model}. It can be implemented explicitly by adding loss terms to the optimization objective, or implicitly through
the design and parameterization of the transformation model itself. 
The following subsections present examples of the most commonly employed explicit analytical regularization formulations.

\textbf{Smoothness.} The most widely implemented regularization in image registration is diffusion regularization \citep{balakrishnan2019voxelmorph,mok2020large}. This constraint penalizes the
first-order spatial derivatives of the displacement field, encouraging smooth and continuous deformations throughout the domain. It minimizes the L2-norm of the gradients of the displacement
field, effectively penalizing disparities between adjacent displacements. Alternatively, curvature regularization minimizes the bending energy by instead penalizing the second-order derivatives of
the displacement field. Another complementary strategy conventionally used in image registration algorithms is multiresolution registration. This strategy both addresses the large deformation
problem and implicitly promotes smoothness, as optimization is constrained to smooth solutions early on, with finer and potentially less smooth details added only at later stages.

\textbf{Topology consistency and volume preservation.} Maintaining the topological integrity of anatomical structures during registration is critical to avoid physical violations by preventing
artifacts such as overlapping regions or implausible tissue compression/expansion. A deformation field $\phi$ is then considered topology-preserving if it is diffeomorphic (i.e.,~smooth and
invertible). This condition is commonly evaluated by analyzing the Jacobian determinant of the deformation field, $|\boldsymbol{J}_\phi|$, at each voxel. Values of $|\boldsymbol{J}_\phi|  > 0$ indicate local invertibility
(i.e.,~no folding or tearing) \citep{haber2007image}, whereas $| \boldsymbol{J}_\phi|  = 1$ implies local volume preservation \citep{rohlfing2003volume}. Thus, negative values of the Jacobian determinant or large
deviations from the unit are penalized to enforce these desirable properties. 

\textbf{Inverse consistency.} Inverse consistency (IC) regularization addresses the symmetry of the registration process, ensuring that the order of registration (i.e.,~whether $I_M$ is
registered to $I_F$ or vice versa) does not introduce directional biases \citep{christensen2002consistent}. This is enforced by penalizing deviations from the identity, $\boldsymbol{Id}$, when
composing forward $\phi_{M,F}$ and backward $\phi_{F,M}$ transformations, so that $\phi_{M,F} \circ \, \phi_{F,M} \simeq \boldsymbol{Id}$.

\begin{figure*}[!t]
    \centering       
    \includegraphics[width=\textwidth,keepaspectratio]{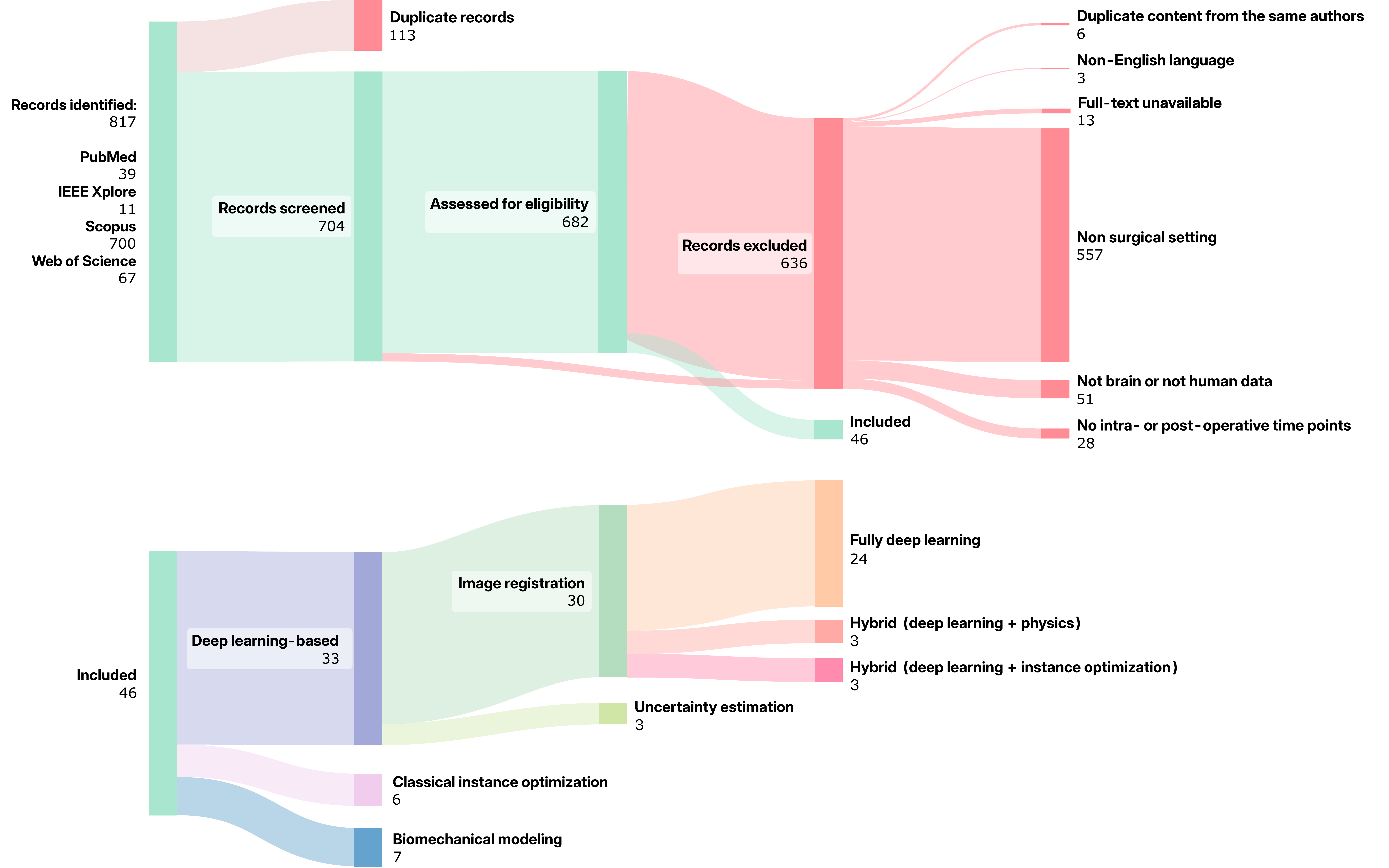}
    \caption{Diagram illustrating the literature search process. Top: a total of 817 records were first identified through database searches in PubMed, IEEE Xplore, Scopus, and Web of Science. Duplicates were removed before screening. Titles and abstracts were reviewed to exclude irrelevant or ineligible studies, followed by full-text assessment based on predefined inclusion criteria. Bottom: the final selection included 46 studies categorized by their methodological approach, including classic instance optimization, biomechanical modeling, and deep learning-based methods.}
    \label{fig:prisma}
\end{figure*}

\section{Methods}
This review applied a systematic search protocol guided by the Preferred Reporting Items for Systematic Reviews and Meta-Analyses (PRISMA) 2020 guidelines \citep{PAGE2021105906}. This section describes the
search strategy employed across multiple scientific databases (Section~\ref{sec:search_strategy}), the multistage screening and article selection process (Section~\ref{sec:screening_process}), the inclusion
and exclusion criteria applied to determine study eligibility (Section~\ref{sec:eligibility_criteria}), and the overall search results and trends observed in the final corpus of studies (Section~\ref{sec:search_results}).

\subsection{Search strategy}
\label{sec:search_strategy}
A comprehensive literature search was performed across four major scientific databases: PubMed, IEEE Xplore, Scopus, and Web of Science. The search was conducted with the publication window restricted to 2020 through 2025 to reflect recent methodological developments in deep learning and data-driven modeling. Only articles published in English and indexed in peer-reviewed venues were considered eligible.

To retrieve articles relevant to the topic of this review, we formulated a query incorporating keywords in brain shift and modeling strategies: \textit{(``brain shift'' OR ``brain deformation'') AND registration AND (model* OR predict* OR simul* OR updat* OR compensat*)}. This final query was iteratively refined to strike a balance between sensitivity and specificity, and was applied uniformly across all databases. The objective was to capture original research that involved modeling, prediction, simulation, or compensation of brain deformation using computational methods, particularly in intraoperative contexts. Supporting citations were also included when necessary to establish context, explain concepts, or discuss observations.

\subsection{Screening process}
\label{sec:screening_process}
The article selection process was carried out in three distinct stages. First, an initial search and de-duplication phase was conducted by one reviewer (TA), who removed duplicate entries and compiled a consolidated list of records. Second, titles and abstracts were screened by five reviewers (TA, CPG, JPC, RD, and IPM) to assess preliminary relevance based on the review objectives. Third, full-text articles of the shortlisted studies were evaluated in detail by four reviewers (TA, CPG, JPC, and NH) to determine their final inclusion.

\begin{figure*}[!t]   
    \centering    
    \includegraphics[width=\textwidth,keepaspectratio]{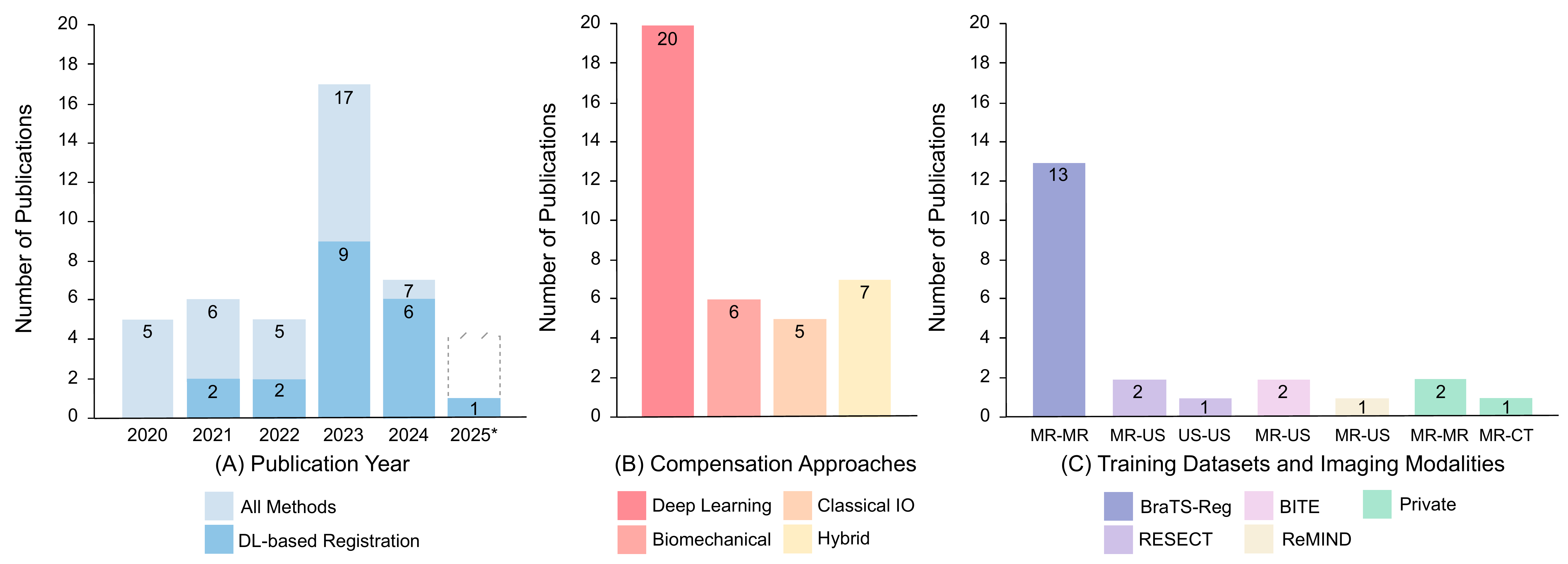}
    \caption{Summary of key trends across studies meeting inclusion criteria. (A) Number of publications per year from 2020 to 2025, comparing all surveyed methods with those exclusively based on DL for image registration. A steady growth trend is observed, with DL approaches dominating recent contributions. (B) Methodological distribution among studies performing image registration. Deep learning-based registration represents the majority, followed by biomechanical modeling and hybrid strategies combining learning with instance optimization (IO) or biomechanics. (C) Registered modality pairs and datasets used in the surveyed studies. While most methods rely on private data, BraTS-Reg is the most widely adopted for magnetic resonance imaging (MRI)-MRI registration, with combined RESECT and BITE data frequently used for MRI-ultrasound (US) and US-US registration to address data scarcity. Surgical microscopy (Micro.), thermography (Therm.), and computed tomography (CT) are the least represented modalities. For a more detailed description of the different public datasets, please refer to Table~\ref{tab:datasets_summary}.}
    \label{fig:stats}
\end{figure*}

\begin{figure}[!t]   
    \centering  
    \includegraphics[width=\columnwidth,keepaspectratio]{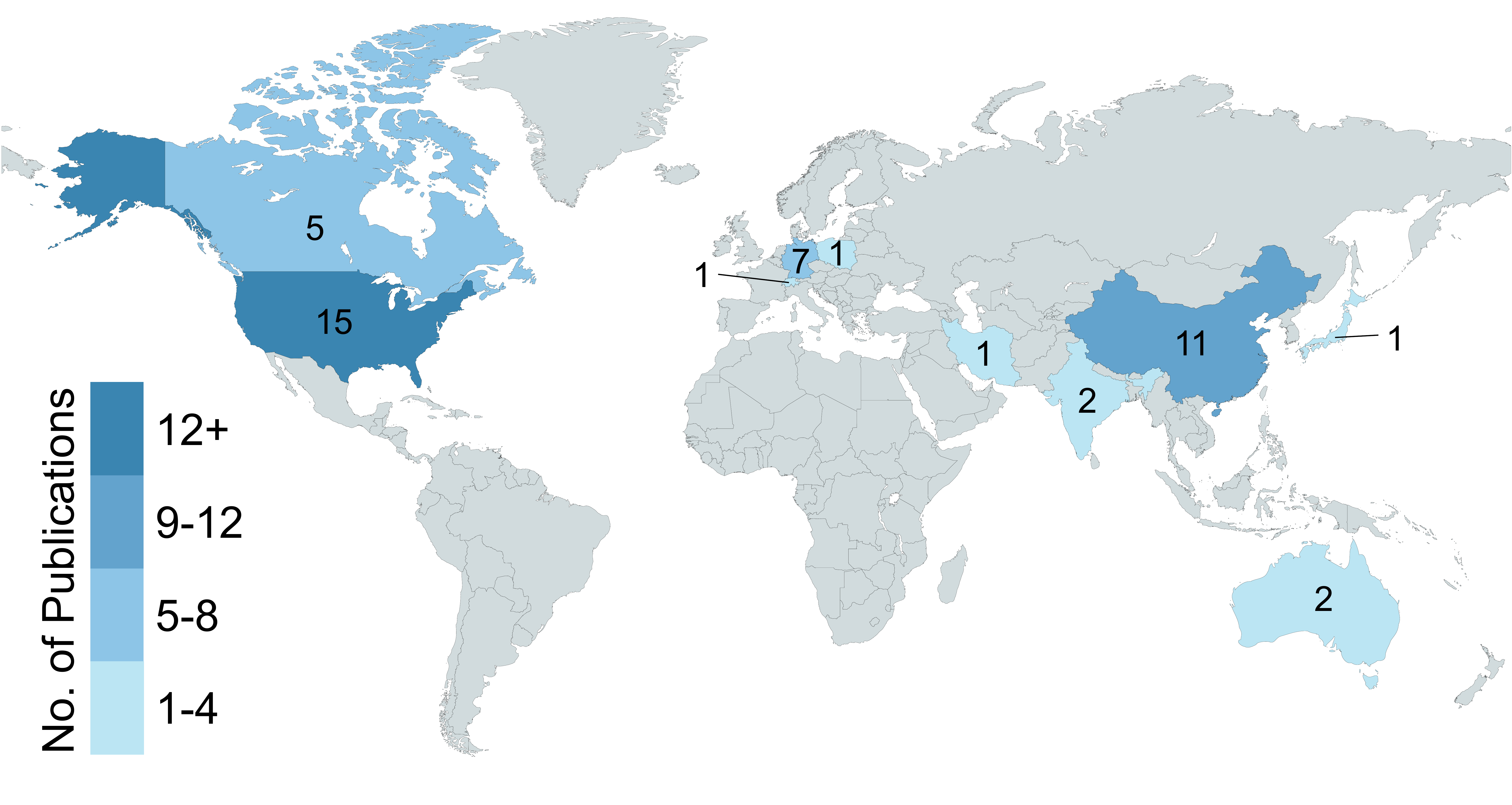}
    \caption{Geographic distribution of the reviewed publications. Darker shading indicates higher publication counts. Research output is dominated by the United States, followed by China and Germany, with additional contributions from Canada and other countries.}
    \label{fig:world_map}
\end{figure}

\subsection{Eligibility criteria}
\label{sec:eligibility_criteria}
Studies were included in the review if they met the following criteria: (1) focused on modeling or compensating for brain deformations between different neurosurgical time points, such as pre- to intra-operative, pre- to post-resection, intra- to post-operative, or pre- to post-operative; (2) utilized imaging modalities relevant to clinical neurosurgery, such as MRI, CT, US, or surgical microscopy; (3) proposed, extended, or indirectly contributed to novel computational methods, including iterative (non-learning), DL-based, or physics-informed approaches, to address the problem of brain shift; and (4) were based on pathological human brain data acquired in a surgical setting. Studies were excluded if they did not satisfy the above conditions. This included research that focused solely on intersubject or atlas-based image registration without modeling deformations directly caused by neurosurgery, employed non-human, non-pathological, or non-brain imaging data, or investigated extended reality not coupled to registration.

\subsection{Search results}
\label{sec:search_results}
The study selection process and its results are summarized in the PRISMA flow diagram in Fig.~\ref{fig:prisma}. An initial search yielded $817$ records across selected databases. After the removal of $113$ duplicates, $704$ unique records remained. Of these, $13$ could not be retrieved in full, three were not in English, and six comprised preliminary results of more complete studies already included. A total of $682$ records were thus screened for eligibility based on title, abstract, and full-text review. Following the application of all inclusion and exclusion criteria, $46$ studies were selected for detailed analysis and inclusion in this review. Studies were primarily excluded due to the absence of neurosurgical deformation modeling, the use of existing techniques without substantive methodological contribution, or a focus unrelated to surgical image registration. Specifically, $26$ studies tackled tasks not influenced by intraoperative changes, such as atlas registration or multimodal fusion in non-surgical contexts, and $51$ did not include human brain data.

\begin{table*}[!t]
\caption{Summary of classic iterative optimization-based registration methods reviewed in this work. Colored squares indicate the dataset(s) used for validation (\raisebox{-.4ex}{\textcolor{red}{\rule{1em}{1em}}}~=~BITE, \raisebox{-.4ex}{\textcolor{green}{\rule{1em}{1em}}}~=~RESECT, \raisebox{-.4ex}{\textcolor{purple}{\rule{1em}{1em}}}~=~BraTS-Reg, \raisebox{-.4ex}{\textcolor{pink}{\rule{1em}{1em}}}~=~Private). Methods are grouped by the registered modalities. Additional details for classic methods can be found in Table~\ref{tab:methods_details}.}
\label{tab:classic_methods_summary}
\resizebox{\textwidth}{!}{
\begin{tabular}{lcc ccccc c ccc c cccc c ccccccc c}
\toprule
\textbf{Method} & \textbf{Dataset} & &
\multicolumn{5}{c}{\textbf{Similarity}} & \multicolumn{1}{c}{} &
\multicolumn{3}{c}{\textbf{Regularization}} & \multicolumn{1}{c}{} &
\multicolumn{4}{c}{\textbf{Optimization}} & \multicolumn{1}{c}{} &
\multicolumn{7}{c}{\textbf{Evaluation}} & \textbf{Code} \\
\cline{4-8} \cline{10-12} \cline{14-17} \cline{19-25}
\addlinespace[4pt]
\multicolumn{1}{c}{} & & &
\multicolumn{1}{c}{\rotatebox{90}{NGF}} &
\multicolumn{1}{c}{\rotatebox{90}{MSE}} &
\multicolumn{1}{c}{\rotatebox{90}{MI}} &
\multicolumn{1}{c}{\rotatebox{90}{NCC}} &
\multicolumn{1}{c}{\rotatebox{90}{SSE}} & &
\multicolumn{1}{c}{\rotatebox{90}{Curvature}} &
\multicolumn{1}{c}{\rotatebox{90}{Jacobian}} &
\multicolumn{1}{c}{\rotatebox{90}{Elastic}} & &
\multicolumn{1}{c}{\rotatebox{90}{L-BFGS}} &
\multicolumn{1}{c}{\rotatebox{90}{PSO}} &
\multicolumn{1}{c}{\rotatebox{90}{CG}} &
\multicolumn{1}{c}{\rotatebox{90}{QPBO}} & &
\multicolumn{1}{c}{\rotatebox{90}{TRE}} &
\multicolumn{1}{c}{\rotatebox{90}{Dice}} &
\multicolumn{1}{c}{\rotatebox{90}{SSIM}} &
\multicolumn{1}{c}{\rotatebox{90}{MAE}} &
\multicolumn{1}{c}{\rotatebox{90}{Robust.}} &
\multicolumn{1}{c}{\rotatebox{90}{MSE}} &
\multicolumn{1}{c}{\rotatebox{90}{PSNR}} & \\
\midrule
\textbf{iUS-to-iUS} & & & & & & & & & & & & & & & & & & & & & & & & & \\
\citet{canalini2020enhanced} & \raisebox{-.6ex}{\textcolor{green}{\rule{1em}{1em}}}/\raisebox{-.6ex}{\textcolor{red}{\rule{1em}{1em}}} & & \yes &  &  &  &  &  & \yes &  &  &  & \yes &  &  &  &  & \yes & \yes &  &  &  &  &  & \xmark \\
\citet{chel2023segmentation} & \raisebox{-.6ex}{\textcolor{green}{\rule{1em}{1em}}}/\raisebox{-.6ex}{\textcolor{red}{\rule{1em}{1em}}} & & & \yes &  &  &  &  &  &  &  &  &  & \yes &  &  &  & \yes &  & \yes &  & \yes &  &  & \xmark \\
\addlinespace[6pt] 
\textbf{preMRI-to-iUS} & & & & & & & & & & & & & & & & & & & & & & & & & \\
\citet{ghose2021automatic} & \raisebox{-.6ex}{\textcolor{green}{\rule{1em}{1em}}} & & &  & \yes & \yes &  &  &  & \yes &  &  &  &  & \yes &  &  & \yes &  &  &  &  &  &  & \xmark \\
\citet{li2026mrf} & \raisebox{-.6ex}{\textcolor{green}{\rule{1em}{1em}}}/\raisebox{-.6ex}{\textcolor{red}{\rule{1em}{1em}}} & & &  &  & \yes &  &  &  & \yes &  &  &  &  &  & \yes &  & \yes &  &  &  &  &  &  & \xmark \\
\addlinespace[6pt] 
\textbf{preMRI-to-postMRI} & & & & & & & & & & & & & & & & & & & & & & & & & \\
\citet{canalini2022iterative} & \raisebox{-.6ex}{\textcolor{purple}{\rule{1em}{1em}}} & & \yes &  &  &  &  &  & \yes & \yes &  &  & \yes &  &  &  &  &  &  &  & \yes & \yes &  &  & \xmark \\
\addlinespace[6pt] 
\textbf{Thermography} & & & & & & & & & & & & & & & & & & & & & & & & & \\
\citet{iorga2023robust} & \raisebox{-.6ex}{\textcolor{pink}{\rule{1em}{1em}}} & & &  &  &  & \yes &  &  &  & \yes &  &  &  & \yes &  &  &  &  & \yes &  &  & \yes & \yes & \xmark \\
\bottomrule
\end{tabular}
}
\vspace{0.1em}
\\ \noindent \footnotesize NGF --- Normalized Gradient Fields; MSE --- Mean Squared Error; MI --- Mutual Information; NCC --- Normalized Cross-Correlation; \hbox{SSE --- Sum of Squared Errors;} TRE --- Target Registration Error; MAE --- Median Absolute Error; SSIM --- Structural Similarity Index Measure; \hbox{L-BFGS --- Limited-memory Broyden--Fletcher--Goldfarb--Shanno;} PSNR --- Peak Signal-to-Noise Ratio;  \hbox{QPBO --- Quadratic Pseudo-Boolean Optimization;} PSO --- Particle Swarm Optimization; CG --- Conjugate Gradient; Robust. --- Robustness.
\end{table*}

Fig.~\ref{fig:stats} summarizes key trends across the included studies. Research on brain deformation modeling grew steadily from 2020, with DL methods emerging as the predominant approach and overall
activity peaking in 2023. This period followed the 2022 release of the Brain Tumor Sequence Registration (BraTS-Reg) dataset and challenge \citep{baheti2024brain}, which enabled standardized
benchmarking and broader model comparison. Among DL-based registration studies, MRI-MRI and MRI-US were the most common modality pairs. BraTS-Reg was the primary dataset for MRI-MRI registration,
followed by the combined use of the REtroSpective Evaluation of Cerebral Tumors (RESECT) \citep{xiao2017retrospective} and Brain Images of Tumors for Evaluation (BITE) \citep{mercier2012online}
datasets, which are frequently combined to address the scarcity of paired MRI-US and US-US data. Several studies relied on private in-house datasets, reflecting the scarcity of public intraoperative
imaging data. Classic optimization-based and biomechanical modeling approaches accounted for a smaller proportion of recent contributions, though hybrid frameworks that combine learning-based
methods with IO or physics-informed modeling have gained traction. Research activity was largely concentrated in the United States, China, Germany, and Canada (Fig.~\ref{fig:world_map}), reflecting established expertise in medical image computing and neurosurgical deformation modeling.

The $46$ studies included in this review were categorized according to their modeling paradigm. Six studies employed classic iterative registration approaches, often refining existing algorithms to improve robustness and accuracy to specific surgical challenges (e.g.,~large deformations, missing correspondences, multimodality). Seven studies implemented biomechanical models simulating intraoperative brain deformation based on patient-specific geometries, tissue properties, and physical constraints. The remaining $33$ studies used DL-based methods, which represented the predominant research direction. Among these, $30$ focused on performing image registration: $24$ relying exclusively on learning-based architectures, three combining DL with IO, and three incorporating biomechanics into the pipeline to enhance biomechanical plausibility. Three studies also estimated registration uncertainty or residual error, enabling probabilistic and interpretable assessments of registration quality.

\section{Image-based registration}
\label{sec:image_registration}
By estimating spatial transformations that align preoperative, intraoperative, and postoperative images, image registration integrates multimodal data and corrects tissue deformations during neurosurgery. Its evolution mirrors the broader shift in medical image computing from handcrafted methods to DL frameworks capable of learning complex deformations directly from data. The following sections outline this progression from classic IO approaches (Section~\ref{sec:classic_instance_optimization}) to learning-based registration (Section~\ref{sec:learning_based_registration}), which encompasses direct displacement field regression (Section~\ref{sec:direct_regression}), feature-based alignment (Section~\ref{sec:landmark_based}), Transformer architectures (Section~\ref{sec:transformer_based}), and synthesis- or adversary-driven multimodal strategies (Section~\ref{sec:gan_synthesis_based}). Further developments, such as multimodality fusion (Section~\ref{sec:multimodality_fusion}) and methods tailored to handle absent correspondences (Section~\ref{sec:missing_correspondences}), have improved robustness and clinical applicability. Finally, hybrid learning frameworks (Section~\ref{sec:hybrid_learning}) combine the adaptability of neural networks with the precision of case-specific optimization.

\subsection{Classic iterative optimization}
\label{sec:classic_instance_optimization}
Classic registration methods that rely on IO have long served as the foundation of computational image alignment in neurosurgery and continue to be widely used in clinical and research settings. However, neurosurgical workflows introduce substantial challenges, which require significant efforts to adapt these traditional formulations. To address this, researchers have extended classic frameworks by introducing strategies to handle non-corresponding regions or bridge modality gaps, as highlighted in the following examples (Table~\ref{tab:classic_methods_summary}).

A central obstacle in longitudinal registration is handling resection cavities, where tissue present in the preoperative scan no longer exists post-resection. \citet{canalini2020enhanced}
addressed this in pre- to post-resection US alignment by segmenting the cavity with a 3D U-Net and excluding it from a variational NGF-driven registration optimized
with a quasi-Newton (L-BFGS) optimizer. This selective masking improved robustness by focusing the optimization solely on corresponding anatomy. In follow-up work on pre- to post-operative MRI
\citep{canalini2022iterative} for the BraTS-Reg challenge \citep{baheti2024brain}, the same registration pipeline was applied without masking, which the authors identified as a major limitation.

Another challenge is preoperative MRI (preMRI)-to-iUS registration, where the differing physical principles create vastly different image appearances. One strategy is to transform the images into
a more comparable domain. \citet{ghose2021automatic} computed Sobel-based ``T1-gradient'' maps to highlight structural boundaries in a representation that is structurally more analogous to iUS,
facilitating standard block-matching registration using NCC and Normalized Mutual Information (NMI). \citet{li2026mrf} approached this problem differently, framing it as a higher-order Markov random field by matching control points
with Gabor texture descriptors built to stay reliable despite ultrasound speckle. A topological constraint was used to prevent the deformation from folding onto itself. The authors further demonstrated
that a MIND-SSC (Self-Similarity Context) rigid pre-alignment is essential to keep the deformable registration stage from collapsing into a local minimum.

For same-modality alignment, \citet{chel2023segmentation} performed iUS-to-iUS registration by segmenting stable hyperechoic structures (e.g.,~falx, choroid plexus) and driving registration only
with these masks, optimizing a robust MSE cost via particle swarm optimization \citep{kennedy1995particle}. Beyond ultrasound, \citet{iorga2023robust} modeled motion in infrared thermography using
a low-dimensional spline field with an elastic regularizer, achieving smooth, real-time corrections evaluated with MSE, Structural Similarity Index Measure (SSIM), and Peak Signal-to-Noise Ratio
(PSNR).

\subsection{Learning-based registration} 
\label{sec:learning_based_registration}
Unlike iterative techniques that solve an optimization problem for each new pair of images, DL models learn an implicit function that can directly infer the deformation field in a single forward
pass. This offers the potential for near-instantaneous registration, a critical advantage in time-sensitive surgical workflows. They are typically trained in an unsupervised or semi-supervised manner,
where the model uses the difference between the deformed moving image and the fixed image to update the network. In contrast, supervised learning-based methods use ground-truth displacements,
landmark correspondences, or anatomical labels as target output during the training process, comparing the network predictions with the ground truths. These ground truths can be estimated by
traditional registration approaches \citep{avants2008symmetric,klein2009elastix} or manually annotated by experts. 

By learning complex feature representations directly from large datasets, these models aim to overcome the generalization limits of relying on handcrafted metrics, promising more robust performance across diverse patient populations and pathologies. Fig.~\ref{fig:dl_archs} illustrates the principal architectural families on which these methods are built, which differ along a few structural axes: how the deformation is represented, ranging from a dense voxel-wise displacement field (Fig.~\ref{fig:dl_archs}-A) to sparse keypoint correspondences (Fig.~\ref{fig:dl_archs}-B); what supplies the training signal, whether a handcrafted similarity loss, a discriminator-learned adversarial criterion (Fig.~\ref{fig:dl_archs}-D), or translation into a shared appearance space (Fig.~\ref{fig:dl_archs}-E); and the receptive field of the backbone, from local convolution to the global self-attention of Transformer-based designs (Fig.~\ref{fig:dl_archs}-C). The following sections describe each family in turn, then cover strategies built on top of them to address specific challenges, namely multimodality fusion, the handling of missing correspondences in longitudinal data, and hybrid frameworks that couple learning-based prediction with iterative optimization. A compilation of DL-based methods for brain deformation compensation addressed in this review is presented in Table~\ref{tab:dl_methods}.

\subsubsection{Direct displacement field regression}
\label{sec:direct_regression}
\begin{figure*}[!t]
    \centering
    \includegraphics[width=1\textwidth, keepaspectratio]{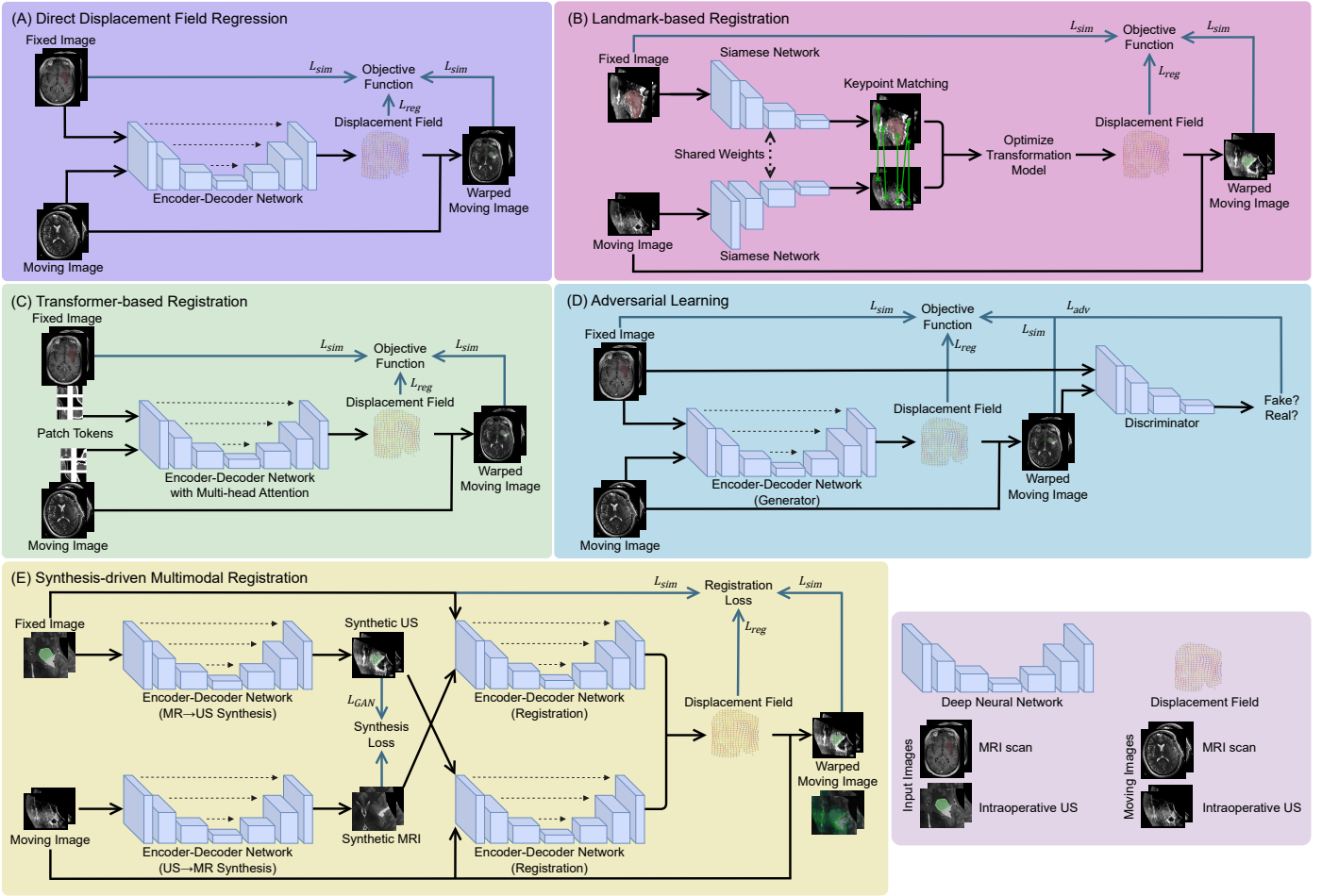}
    \caption{Representative deep learning frameworks for image registration in image-guided neurosurgery. Blue arrows show the inputs to the objective function that guides backpropagation. (A) \textit{Direct Displacement Field Regression} \hbox{(Section~\ref{sec:direct_regression})}: convolutional encoder-decoder networks directly predict dense voxel-wise displacement vectors between moving and fixed images by minimizing a composite loss comprising image similarity ($L_{sim}$) and regularization ($L_{reg}$) terms. (B) \textit{Keypoint-based Registration} \hbox{(Section~\ref{sec:landmark_based})}: these networks detect, match, and/or take sparse anatomical keypoints as input, using these correspondences to derive the displacement field. Green lines and stars represent matched keypoints between images. (C) \textit{Transformer-based Registration} \hbox{(Section~\ref{sec:transformer_based})}: networks employ self-attention mechanisms to capture long-range spatial dependencies and global context, improving structural coherence and alignment across distant anatomical regions. (D) \textit{Adversarial Learning} \hbox{(Section~\ref{sec:gan_synthesis_based})}: registration is guided by a discriminator network that distinguishes between realistic and unrealistic deformations, enforcing anatomical plausibility via a learned similarity measure, or adversarial loss ($L_{adv}$). (E) \textit{Synthesis-driven Multimodal Registration} \hbox{(Section~\ref{sec:gan_synthesis_based})}: modality translation networks (e.g., MRI$\leftrightarrow$US synthesis) generate synthetic counterparts to facilitate cross-modality alignment. Registration and synthesis are jointly optimized with an added synthesis or adversarial loss ($L_{GAN}$).}
    \label{fig:dl_archs}
\end{figure*}

The most common DL formulation for registration directly regresses dense deformation fields using U-Net-like architectures
\citep{de2017end,sokooti2017nonrigid,hu2018weakly,balakrishnan2019voxelmorph} (Fig.~\ref{fig:dl_archs}-A). These models take image pairs as input and directly predict a displacement vector for each
voxel in the image domain. VoxelMorph \citep{balakrishnan2019voxelmorph}, a canonical example for learning-based image registration under this simple formulation, serves as the starting point on which most current methods build and expand.

\begin{table*}[!t]
\caption{Summary of deep learning-based registration methods reviewed in this work. For each method, colored squares indicate the dataset(s) used during training (\raisebox{-.4ex}{\textcolor{red}{\rule{1em}{1em}}}~=~BITE, \raisebox{-.4ex}{\textcolor{green}{\rule{1em}{1em}}}~=~RESECT, \raisebox{-.4ex}{\textcolor{orange}{\rule{1em}{1em}}}~=~ReMIND, \raisebox{-.4ex}{\textcolor{blue}{\rule{1em}{1em}}}~=~BraTS (2021), \raisebox{-.4ex}{\textcolor{purple}{\rule{1em}{1em}}}~=~BraTS-Reg, \raisebox{-.4ex}{\textcolor{pink}{\rule{1em}{1em}}}~=~Private). Methods are grouped by the registered modalities. Additional details for deep learning-based methods can be found in Table~\ref{tab:methods_details}.}
\label{tab:dl_methods}
\resizebox{\textwidth}{!}{
\begin{tabular}{lclclllcclllllccllllllllc}
\toprule
\multicolumn{1}{c}{\textbf{Method}} & \textbf{Dataset} & \multicolumn{5}{c}{\textbf{Similarity Term}} & \multicolumn{1}{c}{} & \multicolumn{6}{c}{\textbf{Regularization Term}} & \multicolumn{1}{c}{} & \multicolumn{9}{c}{\textbf{Evaluation Metric}} & \textbf{Code} \\
\cline{3-7} \cline{9-14} \cline{16-24}
\addlinespace[4pt]
\multicolumn{1}{c}{} & &
\multicolumn{1}{c}{\rotatebox{90}{MSE}} & \rotatebox{90}{NCC} & \multicolumn{1}{c}{\rotatebox{90}{MI}} & \multicolumn{1}{c}{\rotatebox{90}{MIND}} & \multicolumn{1}{c}{\rotatebox{90}{Adversarial}} &
&
\rotatebox{90}{Diffusion} & \multicolumn{1}{c}{\rotatebox{90}{Bending}} & \multicolumn{1}{c}{\rotatebox{90}{Jacobian}} & \multicolumn{1}{c}{\rotatebox{90}{IC}} & \multicolumn{1}{c}{\rotatebox{90}{Edge-map}} & \multicolumn{1}{c}{\rotatebox{90}{Sanity}} &
&
\rotatebox{90}{TRE} & \multicolumn{1}{c}{\rotatebox{90}{Robustness}} & \multicolumn{1}{c}{\rotatebox{90}{MAE}} & \multicolumn{1}{c}{\rotatebox{90}{Dice}} & \multicolumn{1}{c}{\rotatebox{90}{HD}} & \multicolumn{1}{c}{\rotatebox{90}{Jacobian}} & \multicolumn{1}{c}{\rotatebox{90}{RMSE}} & \multicolumn{1}{c}{\rotatebox{90}{CC}} & \multicolumn{1}{c}{\rotatebox{90}{SSD}} & \\
\midrule

\textbf{preMRI-to-iMRI} &  &  &  &  &  &  &  &  &  &  &  &  &  &  &  &  &  &  &  &  &  &  & &   \\
MetaRegNet \citep{joshi2023metaregnet} & \raisebox{-.6ex}{\textcolor{blue}{\rule{1em}{1em}}} & \yes &  &  &  &  &  & \yes &  & \yes & & &  &  &  &  &  & \yes &  & \yes &  &  & \yes & \cmark  \\
\addlinespace[6pt]

\textbf{preMRI-to-iCT} &  &  &  &  &  &  &  &  &  &  &  &  &  &  &  &  &  &  &  &  &  &  & &   \\
\citet{han2022deformable} & \raisebox{-.6ex}{\textcolor{pink}{\rule{1em}{1em}}} &  & \yes &  & \yes & \yes &  & \yes &  &  &  &  &  &  & \yes &  &  & \yes & \yes & \yes &  &  & & \xmark  \\
\addlinespace[6pt]

\textbf{iUS-to-iUS} &  &  &  &  &  &  &  &  &  &  &  &  &  &  &  &  &  &  &  &  &  &  & &   \\
\citet{wodzinski2021adversarial} & \raisebox{-.6ex}{\textcolor{green}{\rule{1em}{1em}}} &  &  &  &  & \yes &  &  &  &  &  &  &  &  & \yes &  &  &  &  &  &  &  &   &\xmark \\
\addlinespace[6pt]

\textbf{preMRI-to-iUS} &  &  &  &  &  &  &  &  &  &  &  &  &  &  &  &  &  &  &  &  &  &  & &   \\
EfficientMorph \citep{aziz2025efficientmorph} & \raisebox{-.6ex}{\textcolor{orange}{\rule{1em}{1em}}} &  & \yes &  &  &  &  &  & \yes &  &  &  &  &  & \yes &  &  &  &  & \yes &  &  & & \cmark  \\
D2BGAN \citep{rahmani2024d2bgan} & \raisebox{-.6ex}{\textcolor{green}{\rule{1em}{1em}}}/\raisebox{-.6ex}{\textcolor{red}{\rule{1em}{1em}}} &  &  & \yes &  & \yes &  &  &  &  &  &  &  &  & \yes &  &  &  &  &  &  &  & & \xmark  \\
iRegNet \citep{zeineldin2021iregnet} & \raisebox{-.6ex}{\textcolor{green}{\rule{1em}{1em}}}/\raisebox{-.6ex}{\textcolor{red}{\rule{1em}{1em}}} & \yes & \yes &  &  &  &  & \yes &  &  &  &  &  &  & \yes &  &  &  &  &  &  &  & & \xmark  \\
SynMSE \citep{zhu2025synmse} & \raisebox{-.6ex}{\textcolor{green}{\rule{1em}{1em}}} &  &  &  & \yes & \yes &  & \yes &  &  &  &  &  &  & \yes &  &  & \yes & \yes & \yes &  &  &  & \cmark  \\
\addlinespace[6pt]

\textbf{preMRI-to-postMRI} &  &  &  &  &  &  &  &  &  &  &  &  &  &  &  &  &  &  &  &  &  &  & &   \\
DIRAC \citep{mok2022unsupervised} & \raisebox{-.6ex}{\textcolor{purple}{\rule{1em}{1em}}} &  & \yes &  &  &  &  & \yes &  &  & \yes &  &  &  & \yes & \yes &  &  &  & \yes &  &  & & \cmark  \\
CRRNet \citep{wu2025crrnet} & \raisebox{-.6ex}{\textcolor{purple}{\rule{1em}{1em}}}/\raisebox{-.6ex}{\textcolor{pink}{\rule{1em}{1em}}} &  & \yes &  &  &  &  & \yes &  &  &  &  &  &  & \yes & \yes &  &  &  & \yes &  &  &  & \xmark \\
MetaLapIRN \citep{pan2025metamorphic} & \raisebox{-.6ex}{\textcolor{purple}{\rule{1em}{1em}}}/\raisebox{-.6ex}{\textcolor{blue}{\rule{1em}{1em}}} &  & \yes &  &  &  &  & \yes &  &  & \yes &  &  &  & \yes & \yes &  &  &  & \yes &  &  &  & \cmark  \\
PULPo \citep{siegert2024pulpo} & \raisebox{-.6ex}{\textcolor{purple}{\rule{1em}{1em}}} &  & \yes &  &  &  &  & \yes &  &  &  &  &  &  & \yes &  &  &  &  & \yes & \yes &  & & \cmark  \\
iRegNet \citep{zeineldin2022self} & \raisebox{-.6ex}{\textcolor{purple}{\rule{1em}{1em}}} &  & \yes &  &  &  &  & \yes &  &  &  &  &  &  & \yes & \yes & \yes &  &  &  &  &  & & \cmark  \\
WSSAMNet \citep{almahfouz2022wssamnet} & \raisebox{-.6ex}{\textcolor{purple}{\rule{1em}{1em}}} &  & \yes & \yes &  &  &  & \yes &  &  &  &  &  &  &  & \yes & \yes &  &  &  &  &  &  & \cmark \\
NICE-Net \citep{meng2022brain} & \raisebox{-.6ex}{\textcolor{purple}{\rule{1em}{1em}}} & \yes & \yes &  &  &  &  & \yes &  & \yes &  &  &  &  &  & \yes & \yes &  &  & \yes &  &  & & \cmark  \\
MSF-AR Net \citep{liu2023multi} & \raisebox{-.6ex}{\textcolor{pink}{\rule{1em}{1em}}} &  & \yes &  &  &  &  & \yes &  & \yes &  &  &  &  &  &  &  &  &  & \yes &  & \yes & & \xmark  \\
\citet{feng2024stepwise} & \raisebox{-.6ex}{\textcolor{purple}{\rule{1em}{1em}}} &  & \yes &  &  &  &  & \yes &  &  & \yes &  &  &  & \yes &  &  &  &  &  &  &  &   & \xmark\\
\citet{zhang2024deep} & \raisebox{-.6ex}{\textcolor{purple}{\rule{1em}{1em}}} &  & \yes &  &  &  &  & \yes &  &  & \yes &  &  &  & \yes & \yes & \yes &  &  & \yes &  &  & & \xmark  \\
\citet{waldmannstetter2023primitive} & \raisebox{-.6ex}{\textcolor{purple}{\rule{1em}{1em}}}/\raisebox{-.6ex}{\textcolor{pink}{\rule{1em}{1em}}} & \yes & \yes &  &  &  &  & \yes &  &  &  &  &  &  & \yes &  &  &  &  &  &  &  & & \xmark  \\
\citet{tang2024deformable} & \raisebox{-.6ex}{\textcolor{purple}{\rule{1em}{1em}}} &  & \yes &  &  &  &  & \yes &  &  & \yes &  &  &  & \yes & \yes &  &  &  & \yes &  &  & & \xmark  \\
\citet{wu2024noise} & \raisebox{-.6ex}{\textcolor{purple}{\rule{1em}{1em}}}/\raisebox{-.6ex}{\textcolor{pink}{\rule{1em}{1em}}} &  & \yes &  &  &  &  & \yes &  &  & \yes &  &  &  & \yes & \yes &  &  &  & \yes &  &  & & \cmark  \\
\citet{abderezaei20223d} & \raisebox{-.6ex}{\textcolor{purple}{\rule{1em}{1em}}} & \yes &  &  &  &  &  & \yes &  &  &  & \yes &  &  &  & \yes & \yes &  &  &  &  &  & & \xmark  \\
\citet{wodzinski2022unsupervised} & \raisebox{-.6ex}{\textcolor{purple}{\rule{1em}{1em}}} &  & \yes &  &  &  &  & \yes &  &  &  &  &  &  &  & \yes & \yes &  &  &  &  &  & & \xmark  \\
\citet{duan2023towards} & \raisebox{-.6ex}{\textcolor{purple}{\rule{1em}{1em}}} &  & \yes &  &  &  &  & \yes &  &  &  &  & \yes &  & \yes  & \yes &  &  &  & \yes & & & & \cmark \\
\bottomrule
\end{tabular}
}

\noindent \\ \footnotesize MSE - Mean Squared Error; NCC - Normalized Cross-Correlation; MI - Mutual Information; MIND - Modality Independent Neighborhood Descriptor; \hbox{IC - Inverse Consistency}; TRE - Target Registration Error; MAE - Median Absolute Error; HD - Hausdorff Distance; RMSE - Root Mean Squared Error; \hbox{CC - Correlation Coefficient}; SSD - Sum of Squared Differences.
\end{table*}

An example is iRegNet by \citet{zeineldin2021iregnet}, originally proposed as a semi-supervised U-Net for preoperative MRI to iUS registration with real-time inference on datasets like BITE and RESECT.
The same backbone was later adapted \citep{zeineldin2022self} to a self-supervised setting for preoperative to follow-up MRI alignment in the BraTS-Reg challenge, illustrating how simple CNN
architectures can be repurposed across modalities and tasks with appropriate loss function design. Supervised strategies have also been used to learn deformation patterns directly from outputs of
classic registration algorithms. For instance, \citet{shimamoto2023precise} trained a dual \hbox{U-Net (W-Net)} to predict brain shift after dural opening using deformation fields generated by the
Demons \citep{thirion1998image} algorithm as supervision. Although it achieved good mean TRE results when compared to standard affine approaches, the upper bound of supervised models is inherently
constrained by the quality and biases of the ground truth used during training.

\subsubsection{Keypoint- and feature-based registration}
\label{sec:landmark_based}
Keypoint- and feature-based methods adopt a different perspective, focusing on sparse salient structures or anatomical label maps to drive the registration rather than relying on dense voxel-wise
alignment \citep{heinrich2022voxelmorph,wang2023robust,rasheed2024learning,wang2024brainmorph} (Fig.~\ref{fig:dl_archs}-B). By detecting and matching keypoints or features across images, these
approaches tend to be more robust to large deformations, partial fields of view, and topological changes. They also offer increased interpretability, as correspondences can be visualized and
quantitatively inspected. Such properties make these methods particularly well-suited for clinical applications involving complex anatomical changes (e.g.,~tumor resection), sparse annotations, or
multimodal data.

\citet{pirhadi2023robust} illustrated this with a two-stage Siamese CNN for iUS-to-iUS registration inspired by object-tracking tasks. Their pipeline uses 2.5D patches across orthogonal
planes to learn correspondences, followed by affine transformation estimation via Iterative Reweighted Least Squares (IRLS) \citep{bergstrom2014robust}. The model was pre-trained on
natural images and fine-tuned on medical data, showing strong performance with lower complexity compared to fully 3D CNNs. However, the approach remains semi-automatic as it relies on manual
annotation of reference landmarks. Moving beyond purely geometric keypoints, \citet{almahfouz2022wssamnet} proposed WSSAMNet, which first segments regions of interest around manually annotated
landmarks and then uses these segmentations as attention maps to guide a subsequent registration step. These maps guide the registration network toward relevant anatomical structures and help
manage scenarios with missing correspondences.

\subsubsection{Transformer-based registration}
\label{sec:transformer_based}
Motivated by the success of Transformers in both natural language processing \citep{vaswani2017attention} and computer vision \citep{dosovitskiy2021an}, several registration methods have explored
attention mechanisms to capture long-range spatial dependencies \citep{chen2022transmorph,shi2022xmorpher} (Fig.~\ref{fig:dl_archs}-C). However, standard 3D Transformer models are computationally
demanding, which limits their practicality for high-resolution neuroimaging.

\citet{aziz2025efficientmorph} addressed this limitation with \hbox{EfficientMorph}, a parameter-efficient Transformer-based model for both uni- and multimodal registration. Their innovations
include plane-wise attention mechanisms, where the model decomposes the problem into a sequence of 2D attentions across orthogonal planes, and efficient high-resolution tokenization by grouping
and concatenating the features of adjacent non-overlapping voxel token blocks. This design significantly reduces memory and parameter count while retaining competitive performance, yielding better
results than other learning-based approaches on the \hbox{ReMIND2Reg} dataset\footnote{\url{https://doi.org/10.5281/zenodo.12700312}}, a preprocessed subset of the Brain Resection Multimodal
Imaging Database (\hbox{ReMIND}) \citep{juvekar2024remind}, with lower TRE, $16\times$ fewer parameters, and $5\times$ faster convergence than TransMorph \citep{chen2022transmorph}. This study
shows that Transformer performance and clinical feasibility are not mutually exclusive, though plane-wise attention may be slightly less effective at capturing complex 3D interactions compared to
true 3D attention.

\subsubsection{Adversarial- and synthesis-driven registration}
\label{sec:gan_synthesis_based}
Multimodal registration, such as MRI-to-US or MRI-to-CT, remains one of the most challenging problems in neurosurgical image analysis due to the large differences in image characteristics across
modalities. Traditional similarity metrics employed in multimodal scenarios, such as NCC or MI, may still fail due to these fundamental differences in intensities, gradients, and noise. To address
this, recent research explored the use of Generative Adversarial Networks (GANs) \citep{goodfellow2020generative} and Multimodal Hierarchical Variational Autoencoders (MHVAEs)
\citep{dorent2025unified}, which either implicitly learn modality-invariant representations through adversarial training (Fig.~\ref{fig:dl_archs}-D) or synthesize cross-domain images into a common
appearance space to facilitate alignment (Fig.~\ref{fig:dl_archs}-E).

The work of \citet{han2022deformable} is an example of a synthesis-driven design that combines a CycleGAN for bidirectional preMRI-iCT translation with a dual-channel registration network that operates on real
preoperative and synthetically deformed intraoperative images. Uncertainty maps learned during synthesis are used to weight the contributions of each modality, ensuring that more reliable
registration paths dominate where confidence is higher. Training incorporated standard cycle-consistency and adversarial losses, along with a novel structure-consistency loss that measures the
L1-norm between MIND features of the synthesized and input images. Similarly, \citet{rahmani2024d2bgan} introduced D2BGAN, a Dual Discriminator Bayesian GAN for preMRI-to-iUS registration
incorporating probabilistic priors in the form of a Bayesian term in the cycle-consistency loss and a standard MI-based loss to enforce cross-modal structural consistency.

Rather than relying solely on explicit similarity losses, some adversarial frameworks use the discriminator itself as a learned surrogate for alignment quality. \citet{wodzinski2021adversarial}
proposed a real-time GAN-based affine registration network for iUS-to-iUS alignment, where the generator directly predicts affine transformation parameters and the discriminator learns to
distinguish aligned from misaligned volume pairs. The model achieved competitive TRE on the RESECT dataset with millisecond inference times. This setup reframes registration as a learned
assessment of alignment quality, allowing the discriminator to encode complex notions of correspondence that are difficult to express with traditional similarity metrics. \citet{zhu2025synmse}
took a complementary route with SynMSE, where a structure-constrained CycleGAN learned the intensity gap between modalities while preserving anatomy through a MIND-based loss, generating paired
images that differ in appearance but share spatial layout. These pairs were used to train a learned evaluator that could score image alignment while ignoring modality-specific intensity differences. SynMSE could then replace handcrafted metrics such as NCC, MI, or MIND in existing registration backbones (demonstrated with VoxelMorph and TransMorph).

Other work, such as that by \citet{dorent2025unified}, used \hbox{MHVAEs} to synthesize MRI from US and demonstrated that synthesis substantially enhances registration accuracy and feature
matching, as the registration then occurs between the same modality. This framework was later leveraged to train a cross-modal feature matcher between US and MRI
\citep{rasheed2024learning,morozov20253d}, further underscoring the value of generative synthesis in reducing cross-modality complexity.

\subsubsection{Multimodality fusion}
\label{sec:multimodality_fusion}
In clinical practice, medical imaging often involves multiple modalities or sequences, each capturing different aspects of brain anatomy or pathology. However, the heterogeneity and anisotropy of these inputs, such as varying spatial resolutions or contrasts, pose significant challenges for learning-based models. Instead of relying on a single input, some novel approaches employ modality fusion strategies that aim to combine data from multiple sources during the registration process.

One example of view-wise fusion is the MSF-AR Net proposed by \citet{liu2023multi}, which addresses the resolution anisotropy commonly found in MRI by processing axial, sagittal, and coronal views
in parallel streams. Its cross-attention-guided fusion module dynamically fuses the outputs of each stream by learning spatial attention weights that select the most trustworthy deformation
information from each plane. \citet{abderezaei20223d} adopted a different strategy, employing an Inception-style \citep{szegedy2015going} network to fuse multiparametric MRI into one channel,
learning and encoding modality-specific features. This fused input is then processed by a TransMorph-based registration network with a Swin Transformer \citep{liu2021swin} encoder for long-range
context modeling. On the BraTS-Reg 2022 dataset, this approach notably outperformed the baseline TransMorph in cases with complex tumor changes. 

\subsubsection{Handling absent correspondences in longitudinal data}
\label{sec:missing_correspondences}
One of the most difficult aspects of brain tumor registration is the presence of non-corresponding regions, such as resection cavities or newly formed lesions, which are present in one image but
not in the other. This violates standard assumptions of smooth and invertible deformations. Failing to account for these changes can lead to implausible deformations, high errors, and distorted
anatomy near the lesion. Recent studies have proposed to tackle this issue by learning to detect and exclude non-corresponding regions during registration and by adopting multiresolution pyramidal
network architectures \citep{mok2020fast,mok2020large,de2019deep,zhao2019recursive}. 

The DIRAC framework by \citet{mok2022unsupervised} is a prominent example that focuses on consistency-based masking. Built on a bidirectional multiresolution backbone (LapIRN
\citep{mok2020large,mok2021conditional}), DIRAC registers image pairs in both forward and backward directions and computes a voxel-wise consistency error. Regions with high inconsistency are
assumed to represent missing or altered anatomy and are excluded from the similarity loss through a learned mask. This prevents the network from forcing unrealistic correspondences near tumor
cavities or resected regions. DIRAC achieved the top rank in the BraTS-Reg 2022 challenge (MICCAI track) and strongly influenced subsequent models. \citet{zhang2024deep} extended this idea to
allow multiparametric MRI integration by incorporating parallel contrast-specific pyramid streams and refining the inconsistency mask using morphological post-processing. This refinement proved
effective in improving the sensitivity and specificity of non-correspondence detection, achieving comparable performance to DIRAC while incorporating richer multimodal cues.

Other recent studies enhance correspondence handling with explicit priors or focused attention. \citet{feng2024stepwise} introduced a stepwise corrected-attention mechanism that modulates network
focus at each resolution level. This mechanism uses the displacement field from previous resolution levels and the corresponding warped segmentation of the preoperative lesion to emphasize stable
tissue while down-weighting pathological areas. This approach improves alignment around resection cavities, but is sensitive to the quality of segmentations. A different strategy is taken by
\citet{meng2022brain}, who presented NICE-Net, a non-iterative cascade network that combines an unsupervised NCC loss with sparse landmark supervision, providing weak geometric guidance in
regions where intensity-based matching is unreliable (i.e.,~the resection cavity). Their model additionally incorporates intersubject pretraining and patient-specific fine-tuning to better adapt to
individual patient anatomy.

Beyond geometry-based masking, \citet{tang2024deformable} proposed an auxiliary-image-based intensity consistency constraint. Instead of relying on consistency calculations between just two
images, the model introduces a third auxiliary image, and compares the direct registration path with an indirect one via the auxiliary scan. Differences between these two paths form a soft weight map that down-weights unreliable
regions. This approach is particularly appealing for longitudinal studies with multiple time points. Similarly, \citet{wu2024noise} proposed a self-explainable approach using internal model
behavior to detect inconsistencies without requiring explicit masks. Their method introduces noise-removed inconsistency activation maps, which highlight regions where the model struggles to
reduce the similarity loss. The resulting maps serve as soft masks to weigh the loss function, improving robustness to unreliable or missing correspondences.

In contrast to masking-based approaches, \citet{joshi2023metaregnet} introduced MetaRegNet, a metamorphic image registration framework that explicitly models both deformation and appearance
changes. Their architecture uses dual branches to learn a smooth spatial deformation and an additive intensity source term representing new or missing structures (e.g.,~resection cavity or tumor
recurrence). Built using Lipschitz continuous residual blocks \citep{joshi2022diffeomorphic,joshi2023r2net}, the model encourages physically plausible deformations while accounting for
lesion-induced mass effects. \citet{pan2025metamorphic} extended the metamorphic idea with MetaLapIRN, using a three-level conditional Laplacian pyramid (cLapIRN \citep{mok2021conditional}) to
handle large pre- to post-resection shifts without the need for affine pre-alignment. Alongside a diffeomorphic velocity field, the network predicts a bounded tumor-specific intensity-variation
field, synthesizing pathological appearance changes rather than masking them. MetaLapIRN achieved lower landmark error and smoother deformations than the masking-based DIRAC on the BraTS-Reg
dataset, particularly near the tumor.

A different approach is proposed by \citet{wu2025crrnet} in CRRNet, where missing correspondences are reconstructed instead. It flags unreliable matches via forward--backward consistency, then
restores them with inpainting-style filling from nearby valid matches plus learned long-range sampling of distant ones, re-estimating the cavity rather than masking it out. The trade-off is that
it falters in large textureless regions, where the correlation cues it depends on are weak.

\subsubsection{Hybrid learning}
\label{sec:hybrid_learning}
Even though DL-based registration models enable fast inference, they may fail to capture patient-specific details or generalize to out-of-distribution (OOD) data. To mitigate this, hybrid registration frameworks have been proposed that integrate classic IO with learning-based models at different stages of the pipeline. In these approaches, IO may be used to provide robust initial alignment, to refine deep network predictions at test time, or to be interleaved with learning-based components across multiple stages. In doing so, these frameworks iteratively tailor the deformations to the patient's specific anatomy and pathology.

A key example is the work of \citet{wodzinski2022unsupervised}, who proposed a multistage pipeline that first performs affine registration via IO, followed by nonrigid deformation prediction using
a modified LapIRN. The predicted displacement field is then refined once again through a multiresolution nonrigid optimization stage guided by IC. During this refinement, voxel-wise weights
derived from forward--backward consistency modulate each term of the objective function, enabling precise corrections in challenging regions. This hybrid strategy won first place in the BraTS-Reg
2022 challenge (IEEE ISBI track), highlighting the benefits of combining global, data-driven predictions with local, optimization-based refinements. Similarly, \citet{waldmannstetter2023primitive}
extended VoxelMorph with a test-time IO refinement phase, showing systematic improvements near tumor margins. The model was trained with a composite MSE and NCC loss to balance global intensity
consistency with local structural alignment. Applied to preMRI-to-iMRI registration, this step consistently improved TRE over baseline VoxelMorph. The study also demonstrated that combining
complementary similarity metrics improves convergence and robustness, suggesting hybrid loss designs can outperform single-metric formulations. In multimodal registration, \citet{ha2021modality}
proposed a modality-agnostic, self-supervised framework that pairs deep feature learning with classic optimization for \hbox{MRI-iUS} registration. A 3D CNN is trained to predict MIND-SSC
descriptors, producing dense, modality-invariant features used to estimate probabilistic displacements on a sparse control grid, which are then interpolated and refined with a fast IO
postprocessing step. 

As the field matures, these two components are no longer viewed as mutually exclusive but rather as synergistic tools: deep learning provides powerful priors and initial estimates while classic optimization remains valuable for robustness, especially in complex and unique pathological settings.

\section{Biomechanical modeling and physics-guided registration}
\label{sec:biomech_modeling_physics_registration}

\begin{figure}[!t]   
    \centering  
    \includegraphics[width=\linewidth]{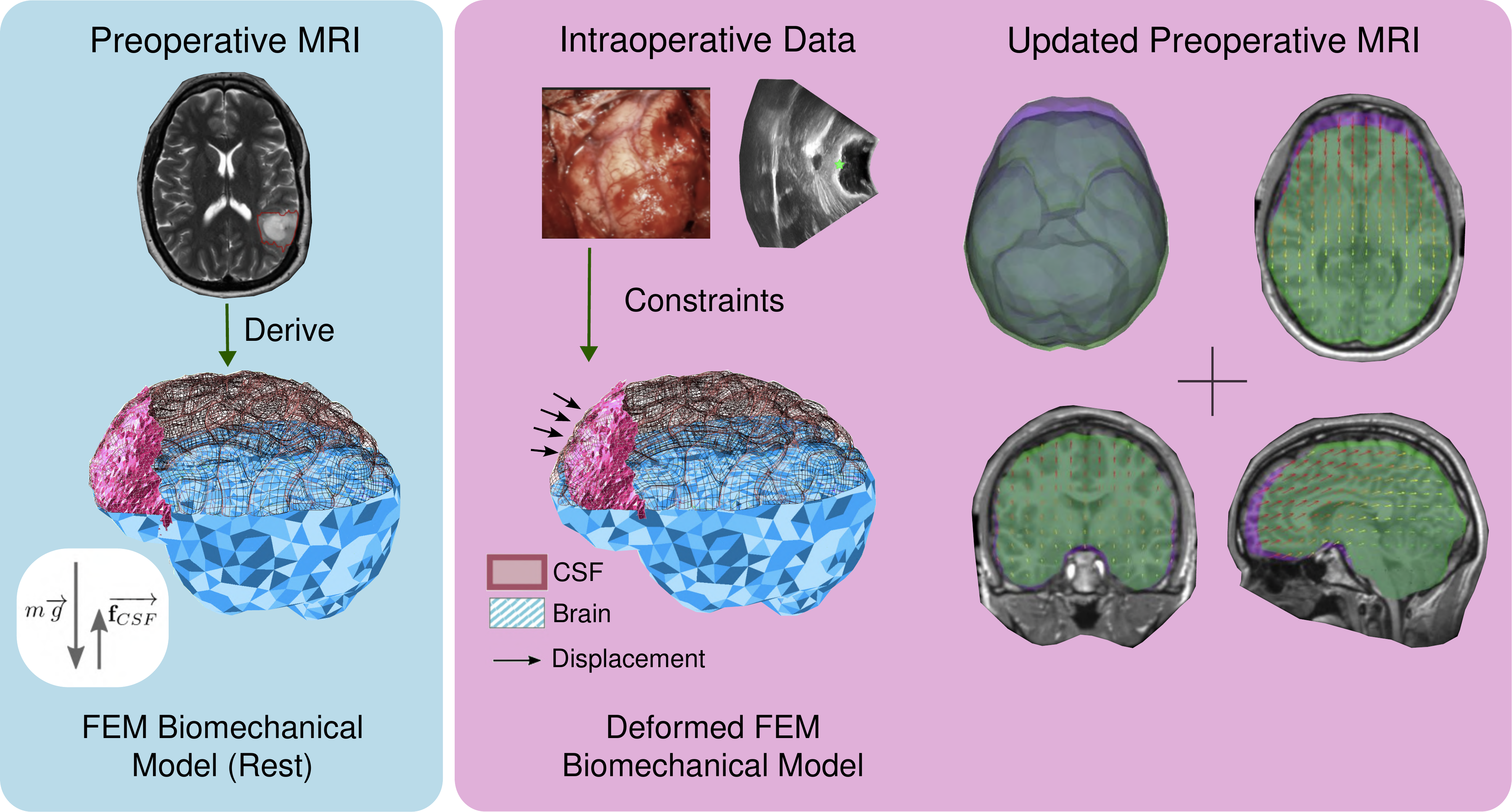}
    \caption{Overview of the biomechanical modeling workflow for image-guided surgery. The pipeline begins with discretizing the anatomical domain from preoperative images into a finite element model (FEM) (in blue; image courtesy of \citet{bilger2011biomechanical}) defined by geometry, material properties, loading forces, and boundary conditions governed by physical laws. Tissue displacement is then computed under intraoperative constraints (e.g.,~ultrasound, surgical imaging, or cerebrospinal fluid amount), and the preoperative images are subsequently updated to reflect the resulting deformations.}
    \label{fig:biomechanical_models}
\end{figure}

Researchers have developed a wide range of biomechanical strategies to model brain deformation during neurosurgery
\citep{mostayed2013biomechanical,miller2019biomechanical,drakopoulos2021adaptive,haouchine2021predicted,yu2022automatic}. These approaches typically share similar use of preoperative imaging but
diverge in how they incorporate intraoperative data. This divergence is dictated by the type and amount of intraoperative information available, ranging from volumetric modalities (e.g.,~MRI, US)
to partial data (e.g.,~cameras, laser-based systems) and physiological or procedural cues (e.g.,~CSF drainage volume, head orientation), which act as loads or constraints on the biomechanical
model. Fig.~\ref{fig:biomechanical_models} illustrates the traditional biomechanical modeling pipeline. The process begins by discretizing the anatomical domain, allowing for the formulation of a
stiffness matrix. Given simulated forces, tissue displacement is computed, after which the biomechanical model is updated. Below, we first outline the foundational principles of biomechanical
modeling and their use in intraoperative registration, then recent DL-based and physics-guided alternatives. Section~\ref{sec:bio_constrained} reviews biomechanical models for intraoperative
registration, and Section~\ref{sec:DL_physics} discusses DL techniques for physics-based brain modeling and physics-guided registration.

\subsection{Biomechanical modeling}
\label{sec:bio_constrained}
Constructing a biomechanical model of the brain can serve multiple purposes: surgical training \citep{talbot2015surgery}, preoperative planning \citep{frisken2021incorporating}, or intraoperative
registration \citep{mostayed2013biomechanical,morin2017brain,luo2020accounting}. Here, our primary interest lies in intraoperative registration or re-registration.

Incorporating biomechanical models in intraoperative registration has been explored for decades. The methods reviewed here are summarized in Table~\ref{tab:physics_methods}. When intraoperative
imaging is available, particularly MRI \citep{luo2020accounting} or US \citep{morin2017brain}, biomechanical models have been used to regularize the registration, enhancing plausibility and
robustness. Their accuracy is comparable to geometric B-spline or thin-plate spline (TPS) approaches \citep{mostayed2013biomechanical,frisken2020comparison}, with a robustness advantage under
large deformations or sparse data. Patient-specific biphasic modeling has been applied by \citet{luo2020accounting}, where a deformation atlas accounts for gravity and CSF loss, and the
simulations are driven by sparse landmark displacements from iMRI to correct shift. For tumor resection, \citet{drakopoulos2021adaptive} proposed an Adaptive Physics-Based Non-Rigid Registration framework in which a heterogeneous linear elastic model is iteratively modified by removing mesh elements in the resected region, capturing soft-tissue deformations near the tumor
margin. The previously mentioned relationship to spline-based alternatives was examined directly by \citet{frisken2020comparison}. In their study, a poroelastic FEM and a TPS driven by the same automatically matched iUS keypoints gave TRE differences that were not statistically significant across the cohort. The FEM, however, produced more consistent deformations away from the driving features.

In the absence of complete intraoperative imaging, biomechanical models can incorporate sparse intraoperative cues such as head orientation, gravity-induced shift, or CSF loss to update the
preoperative anatomy \citep{dumpuri2010fast,chen2013integrating,haouchine2021predicted}. Their role then shifts from registration to prediction, which requires more advanced modeling and tissue
characterization. \citet{lesage2021viscoelastic} explored this context with a linear viscoelastic FEM conditioned on head orientation and CSF drainage, achieving favorable results in low-grade
glioma cases. \citet{narasimhan2020accounting} extended an atlas-based framework with a debulking atlas that simulates tumor growth followed by resection to capture the global mass effect of
cavity collapse, driven by sparse cortical points within a poroelastic FEM. \citet{yu2022automatic} took a different direction with a meshless implementation built on the Meshless Total Lagrangian
Explicit Dynamics (MTLED) algorithm \citep{horton2010meshless}, which replaces the conformal mesh with a point cloud, coupled with an Ogden hyperelastic law and conditioned only on head
orientation and preoperative tumor geometry. Further methods have also been proposed that leverage partial volumetric data or surface information, using manually delineated cortical surfaces
\citep{luo2020accounting}, intraoperative stereovision (iSV) \citep{li2026fully}, laser range scanning \citep{zhuang2011sparse,sun2014near}, or 2D microscope images
\citep{haouchine2021pose,haouchine2023learning}. In such scenarios, biomechanical models act as a regularization prior where data are observed, and play a predictive role in estimating the
unobserved deformation field.

\begin{table*}[!t]
\caption{Summary of biomechanical modeling methods for intraoperative brain deformation compensation reviewed in this work. For each method, colored squares indicate the dataset used for validation (\raisebox{-.4ex}{\textcolor{green}{\rule{1em}{1em}}}~=~RESECT, \raisebox{-.4ex}{\textcolor{pink}{\rule{1em}{1em}}}~=~Private). Methods are grouped by their spatial discretization strategy (finite element or meshless). Additional details on biomechanical modeling methods can be found in Table~\ref{tab:methods_details}.} 
\label{tab:physics_methods}
\fontsize{8}{10}\selectfont
\begin{tabularx}{\textwidth}{p{2.9cm}p{0.8cm}p{3.2cm}p{5.6cm}p{2.7cm}c}
\toprule
\textbf{Method} & \textbf{Dataset} & \textbf{Physical Model} & \textbf{Intraoperative Input} & \textbf{Evaluation Metric} & \textbf{Code} \\
\midrule
\textbf{FEM} &  &  &  & &  \\
\citet{li2026fully} & \hspace{0.25cm} \raisebox{-.6ex}{\textcolor{pink}{\rule{1em}{1em}}} & Linear elastic & Sparse landmark displacements (iSV) & StSD, TRE, PSD & \xmark \\
\citet{frisken2020comparison} & \hspace{0.25cm} \raisebox{-.6ex}{\textcolor{pink}{\rule{1em}{1em}}} & Poroelastic & Sparse landmark displacements (iUS) & TRE & \xmark \\
\citet{luo2020accounting} & \hspace{0.25cm} \raisebox{-.6ex}{\textcolor{pink}{\rule{1em}{1em}}} & Biphasic hyperelastic & Sparse landmark displacements (iMRI) & Euclidean distance, \newline Percent correction & \xmark \\
\citet{narasimhan2020accounting} & \hspace{0.25cm} \raisebox{-.6ex}{\textcolor{pink}{\rule{1em}{1em}}} & Poroelastic & Sparse landmark displacements (LRS, iMRI, iUS) & Euclidean distance, \newline Percent correction & \xmark \\
\citet{lesage2021viscoelastic} & \hspace{0.25cm} \raisebox{-.6ex}{\textcolor{green}{\rule{1em}{1em}}} & Zener linear viscoelastic & Head orientation, CSF loss & TRE & \xmark \\
\citet{drakopoulos2021adaptive} & \hspace{0.25cm} \raisebox{-.6ex}{\textcolor{pink}{\rule{1em}{1em}}} & Linear elastic & Sparse landmark displacements (iMRI) & HD, TRE & \xmark \\

\addlinespace[6pt]
\textbf{Meshless} &  &  &  &  &  \\
\citet{yu2022automatic} & \hspace{0.25cm} \raisebox{-.6ex}{\textcolor{pink}{\rule{1em}{1em}}} & Ogden hyperelastic & Head orientation & HD, Euclidean distance & \cmark$\,^a$ \\
\bottomrule
\end{tabularx}
\vspace{0.5pt} \\
\footnotesize FEM --- Finite Element Method; LRS --- Laser Range Scanner; CSF --- Cerebrospinal Fluid; TRE --- Target Registration Error; StSD --- Stylus-to-Surface Distance; PSD --- Point-to-Surface Distance; HD --- Hausdorff Distance. \\ 
$^a$ Code made available as part of the SlicerCBM extension \citep{safdar2023slicercbm}.
\end{table*}

\begin{table*}[!t]
\caption{Summary of deep learning methods incorporating physics-informed priors for brain deformation modeling reviewed in this work. For each method, colored squares indicate the dataset used during training (\raisebox{-.4ex}{\textcolor{pink}{\rule{1em}{1em}}}~=~Private). Methods are grouped by the registered modalities. Additional details on physics-informed deep learning methods can be found in Table~\ref{tab:methods_details}.} 
\label{tab:dl_physics_hybrid_methods}
\fontsize{8}{10}\selectfont
\begin{tabularx}{\textwidth}{p{4.0cm}p{1.0cm}p{2.8cm}p{2.8cm}p{1.9cm}p{2.2cm}c}
\toprule
\textbf{Method} & \textbf{Dataset} & \textbf{Physics Incorporation \newline Strategy} & \textbf{Loss} & \textbf{Regularization} & \textbf{Evaluation \newline Metric} & \textbf{Code} \\
\midrule
\textbf{preMRI-to-iMRI} &  &  &  &  &  &  \\
PhysGNN \newline \citep{salehi2022physgnn} & \hspace{0.25cm} \raisebox{-.6ex}{\textcolor{pink}{\rule{1em}{1em}}} & FEM simulations of displaced nodes for network supervision & MEE & N/A & MAE, MEE  & \cmark \\

\addlinespace[6pt]
\textbf{Surgical Microscopy} &  &  &  &   &  &  \\
\citet{haouchine2021predicted} & \hspace{0.25cm} \raisebox{-.6ex}{\textcolor{pink}{\rule{1em}{1em}}} & FEM simulations of displaced cortical surface labels for image analogy synthesis & Style error & Smoothness & Mean cortical shift, Dice score  & \xmark \\
\addlinespace[6pt]
\citet{haouchine2021pose} & \hspace{0.25cm} \raisebox{-.6ex}{\textcolor{pink}{\rule{1em}{1em}}} & Shape-from-template deformation modeling of cortical vessels & BCE, Reprojection error & Internal forces \newline priors & MEE, TRE  & \xmark \\
\bottomrule
\end{tabularx}
\vspace{0.5pt} \\
\footnotesize FEM --- Finite Element Method; BCE --- Binary Cross-Entropy; MAE --- Mean Absolute Error; MEE --- Mean Euclidean Error;  TRE --- Target Registration Error.
\end{table*}

\subsection{Deep learning and physics-guided registration}
\label{sec:DL_physics}
Due to the high computational cost of FEM-based simulations, numerous strategies have been proposed to improve efficiency, including domain decomposition, parallel computing, adaptive meshing, and
model order reduction. More recently, DL approaches have emerged, inspired by PINNs, which embed PDEs directly into the training process. These models enable learning from limited data while
enforcing physical consistency, with applications in image-guided neurosurgery (IGN) \citep{tonutti2017machine,salehi2022physgnn}.

Early DL-based approaches relied on simple U-Net architectures to learn a nonlinear mapping from BCs to displacement fields generated by FEM simulations, using standard supervised losses such as
MSE \citep{pfeiffer2019learning,mendizabal2020simulation}. To improve the physical plausibility of predictions, subsequent methods introduced physics-informed penalty terms into the loss function.
For instance, DeepPhysics \citep{odot2022deepphysics} incorporates the residuals of force equilibrium, while \citet{alvarez2024deformable} proposed penalizing violations of force balance by
minimizing the divergence of the Piola--Kirchhoff stress tensor, effectively enforcing biomechanical equilibrium under natural BCs.
More recent efforts have focused on improving model generalizability to unseen geometries, BCs, and material properties. Hypernetworks have been employed to condition deformation models on varying
tissue properties and BCs \citep{el2024hyperu,reithmeir2024data}, while Z-score normalization has been proposed to adapt to diverse mesh topologies \citep{hu2025real}. Graph-based models such as
PhysGNN \citep{salehi2022physgnn} approximate FEM behavior by propagating structural and nodal features across brain meshes, capturing local and global tissue interactions while keeping the number of learnable parameters independent of mesh resolution. Similarly, \citet{min2024biomechanics} proposed a PointNet-based model to achieve geometry- and material-agnostic representations, further improving flexibility and patient specificity.

Formally, the contribution of biomechanical modeling to registration can be expressed through the formulation of BCs, typically categorized as Dirichlet (prescribed displacements) or Neumann
(applied forces or pressures). Dirichlet BCs constrain displacement at specific nodes or surfaces (e.g.,~fixing the brainstem, or enforcing measured displacements), whereas Neumann BCs represent
external loads such as gravity or CSF hydrostatic pressure. With PINNs, this distinction is only partially preserved through loss function design. Dirichlet conditions are typically enforced via
supervised losses, directly regressing predicted displacements to known values, whereas Neumann conditions require differentiating the network output to obtain stress or strain fields and are
imposed by penalizing deviations from equilibrium. The physics-informed DL methods for brain deformation reviewed here are summarized in Table~\ref{tab:dl_physics_hybrid_methods}. For brain registration, \citet{amiri2025physics} combined a predictor step minimizing image dissimilarity with a correction step enforcing
physical equilibrium under elasticity, formalized as a Dirichlet problem. PhysGNN \citep{salehi2022physgnn} approximates FEM-based brain deformation directly by learning from mesh-structured data,
where BCs are implicitly encoded via node connectivity and spatial relationships, enabling fast and accurate registration. In the surgical microscopy setting, \citet{haouchine2021predicted} used
FEM simulations of cortical surface deformations driven by gravity and CSF loss to synthesize image analogies of the expected intraoperative appearance via texture transfer, providing predicted
views of the brain surface before the skull is opened. Pose estimation and nonrigid registration from 2D microscope images were tackled by \citet{haouchine2021pose} as a shape-from-template
problem on cortical vessels, regularized by a wire-like beam model that connects 3D nodes along vessel centerlines and propagates surface displacements to subcortical structures in real time.

\section{Registration uncertainty} 
Most registration models produce a single deformation and rarely communicate how reliable that prediction is. This poses a critical limitation for clinical applications in neurosurgery. In
practice, residual misalignment is inevitable, especially in the presence of pathology, modality disparities, or image artifacts. This has led to a growing interest in modeling not just the
deformation, but also its uncertainty and expected error \citep{geshvadi2025optimizing}. There are two main approaches: estimating the residual misalignment after registration, or embedding
probabilistic uncertainty into the model's prediction process.

Manually assessing the quality of MRI-US registration is difficult and prone to error due to the inherent complexity of 3D surgical data, time constraints in the OR, and the differing contrasts
between MRI and US images. Dense error estimation has emerged as a means to evaluate both linear and nonlinear registration outcomes and provide actionable visual feedback.
\citet{bierbrier2023toward} pioneered a dense error estimation framework for multimodal registration by adapting a sliding-window CNN, originally proposed by \citet{eppenhof2018error}, to predict
voxel-wise registration errors. Training was performed using simulated US volumes generated from preoperative MRI, enabling supervised learning with known synthetic misalignments. Building on this, \citet{salari2023dense} introduced a Swin UNETR-based architecture capable of predicting dense 3D error maps from preMRI-iUS image pairs. By leveraging the global self-attention
capabilities of Swin Transformers, the model captured long-range dependencies critical for multimodal alignment assessment and achieved strong performance on real intraoperative data. Extending
this framework, \citet{salari2023focalerrornet} proposed \hbox{FocalErrorNet}, a 3D focal modulation network that not only estimates patch-wise registration error but also quantifies
uncertainty via Monte Carlo dropout \citep{gal2016dropout,gal2017concrete}. This dual prediction framework provided reliable estimates of both the magnitude and confidence of misalignment, marking
the first such approach in multimodal registration error assessment.

Although these methods focus on \textit{post hoc} estimation of error and uncertainty, another promising line of work aims to embed uncertainty and sanity mechanisms directly into the registration
model itself, resulting in inherently interpretable predictions. One such direction is the work by \citet{siegert2024pulpo}, who proposed PULPo, a probabilistic registration framework built upon a
conditional variational autoencoder (cVAE). PULPo decomposes the deformation field into a hierarchy of stationary velocity fields, arranged in a Laplacian pyramid across multiple spatial scales.
At each level, the model predicts not only a mean deformation but also a variance, enabling voxel-wise uncertainty estimation. The final deformation is sampled from the learned distributions,
allowing the generation of uncertainty maps that reflect both anatomical stability and prediction confidence. Its hierarchical design also enables it to capture uncertainty at both coarse and fine
scales, which is important in regions with both global shifts and local pathology-induced disruptions.

These works reflect a growing recognition that, beyond performance, a model must be interpretable, stable, and clinically trustworthy, particularly when used in sensitive applications such as neurosurgical guidance. These points are further discussed in Section~\ref{sec:disc_clinical}.

\section{Validation strategies and benchmarking}
\label{sec:validation}

\begin{table}[!t]
\begin{centering}
\caption{Registration error and runtime as reported per method, evaluated on public benchmarks and grouped by dataset, registration task, and method class. Values may not be directly comparable across rows, as errors depend on the dataset, task, and landmark protocol, while runtimes depend on the hardware and pipeline. Experimental settings are detailed per-study in Table~\ref{tab:methods_details}.}
\label{tab:registration_results}
\fontsize{8.2}{10.25}\selectfont
\begin{tabular}{lcc}
\toprule
\textbf{Method} & \textbf{\begin{tabular}[c]{@{}c@{}}Registration\\ Error\\ (mm)\end{tabular}} & \textbf{\begin{tabular}[c]{@{}c@{}}Runtime\\ (s)\end{tabular}} \\
\midrule
\multicolumn{3}{l}{\raisebox{-.3ex}{\textcolor{green}{\rule{1em}{1em}}} \textbf{RESECT -- iUS-to-iUS}} \\
\multicolumn{3}{l}{\textit{Classic iterative optimization}} \\
\hspace{1em}\citet{canalini2020enhanced} & 1.21$_{\pm0.66}$ & 5--55 $^a$ \\
\hspace{1em}\citet{chel2023segmentation} & 1.33 & N/A \\
\multicolumn{3}{l}{\textit{Deep learning}} \\
\hspace{1em}\citet{wodzinski2021adversarial} & 1.51$_{\pm0.48}$ & 0.042 \\
\hspace{1em}\citet{pirhadi2023robust} & 1.22$_{\pm0.46}$ & 50 $^b$ \\
\addlinespace[4pt]
\multicolumn{3}{l}{\raisebox{-.3ex}{\textcolor{green}{\rule{1em}{1em}}} \textbf{RESECT -- preMRI-to-iUS}} \\
\multicolumn{3}{l}{\textit{Classic iterative optimization}} \\
\hspace{1em}\citet{ghose2021automatic} & 4.89$_{\pm4.27}$ & N/A \\
\hspace{1em}\citet{li2026mrf} & 1.72$_{\pm0.39}$ & 104.28 \\
\multicolumn{3}{l}{\textit{Deep learning}} \\
\hspace{1em}D2BGAN \citep{rahmani2024d2bgan} & 0.76$_{\pm0.30}$ & N/A \\
\hspace{1em}iRegNet \citep{zeineldin2021iregnet} & 0.84$_{\pm0.16}$ & 0.5 \\
\hspace{1em}\citet{ha2021modality} & 2.33$_{\pm1.70}$ & 3 \\
\addlinespace[4pt]
\multicolumn{3}{l}{\raisebox{-.3ex}{\textcolor{red}{\rule{1em}{1em}}} \textbf{BITE -- iUS-to-iUS}} \\
\multicolumn{3}{l}{\textit{Classic iterative optimization}} \\
\hspace{1em}\citet{canalini2020enhanced} & 2.38$_{\pm2.78}$ & 5--55 $^a$ \\
\hspace{1em}\citet{chel2023segmentation} & 3.67 & N/A \\
\multicolumn{3}{l}{\textit{Deep learning}} \\
\hspace{1em}\citet{pirhadi2023robust} & 1.76$_{\pm1.48}$ & 50 $^b$ \\
\addlinespace[4pt]
\multicolumn{3}{l}{\raisebox{-.3ex}{\textcolor{red}{\rule{1em}{1em}}} \textbf{BITE -- preMRI-to-iUS}} \\
\multicolumn{3}{l}{\textit{Classic iterative optimization}} \\
\hspace{1em}\citet{li2026mrf} & 1.38 & 104.28 \\
\multicolumn{3}{l}{\textit{Deep learning}} \\
\hspace{1em}D2BGAN \citep{rahmani2024d2bgan} & 1.35$_{\pm0.49}$ & N/A \\
\hspace{1em}iRegNet \citep{zeineldin2021iregnet} & 1.47$_{\pm0.61}$ & 0.5 \\
\addlinespace[4pt]
\multicolumn{3}{l}{\raisebox{-.3ex}{\textcolor{orange}{\rule{1em}{1em}}} \textbf{ReMIND -- preMRI-to-iUS}} \\
\multicolumn{3}{l}{\textit{Deep learning}} \\
\hspace{1em}EfficientMorph \citep{aziz2025efficientmorph} & 3.60$_{\pm0.62}$ & N/A \\
\addlinespace[4pt]
\multicolumn{3}{l}{\raisebox{-.3ex}{\textcolor{purple}{\rule{1em}{1em}}} \textbf{BraTS-Reg -- preMRI-to-postMRI}} \\
\multicolumn{3}{l}{\textit{Classic iterative optimization}} \\
\hspace{1em}\citet{canalini2022iterative} & 1.98$_{\pm0.77}$ & N/A \\
\multicolumn{3}{l}{\textit{Deep learning}} \\
\hspace{1em}DIRAC \citep{mok2022unsupervised}$^{\text{NT}}$ & 3.26$_{\pm2.78}$ & 0.023$_{\pm0.005}$ \\
\hspace{1em}DIRAC \citep{mok2022unsupervised}$^{\text{FT}}$ & 1.86$_{\pm0.98}$ & 0.023$_{\pm0.005}$ \\
\hspace{1em}CRRNet \citep{wu2025crrnet}$^{\text{NT}}$ & 2.92$_{\pm2.37}$ & N/A \\
\hspace{1em}CRRNet \citep{wu2025crrnet}$^{\text{FT}}$ & 1.70$_{\pm0.76}$ & N/A \\
\hspace{1em}MetaLapIRN \citep{pan2025metamorphic} & 2.41$_{\pm0.42}$ & 0.440$_{\pm0.003}$ \\
\hspace{1em}iRegNet \citep{zeineldin2022self} & 2.93$_{\pm0.84}$ & 1 \\
\hspace{1em}NICE-Net \citep{meng2022brain} & 3.39 & N/A \\
\hspace{1em}\citet{wu2024noise}$^{\text{NT}}$ & 3.18$_{\pm2.45}$ & N/A \\
\hspace{1em}\citet{wu2024noise}$^{\text{FT}}$ & 1.77$_{\pm0.80}$ & N/A \\
\hspace{1em}\citet{feng2024stepwise} & 1.85$_{\pm0.83}$ & N/A \\
\hspace{1em}\citet{zhang2024deep} & 1.82$_{\pm0.94}$ & N/A \\
\hspace{1em}\citet{tang2024deformable} & 2.38$_{\pm1.19}$ & N/A \\
\hspace{1em}\citet{abderezaei20223d} & 2.91 & N/A \\
\hspace{1em}\citet{wodzinski2022unsupervised} & 1.71$_{\pm0.86}$ & 57.9 $^c$ \\
\hspace{1em}\citet{duan2023towards} & 2.72$_{\pm0.26}$ & N/A \\
\bottomrule
\end{tabular}
\end{centering}
\vspace{0.2em}
\\
\footnotesize
NT --- Near tumor ($\leq 30$ mm); FT --- Far from tumor ($>30$ mm); \\
$^a$ Range for inference only vs. full pipeline including segmentation; \\ 
$^b$ Includes keypoint detection; \\ 
$^c$ Includes \textit{post hoc} instance optimization.
\end{table}

Robust validation is essential for assessing the effectiveness, reliability, and clinical readiness of DL-based brain registration methods. However, the diversity of modalities, clinical settings, and task objectives has resulted in a wide range of evaluation strategies being employed, making direct comparison across studies difficult. In this section, we summarize how validation has been approached in the literature and outline key benchmarking practices.

The studies reviewed in this manuscript employ a mixture of real clinical datasets and synthetic data to validate registration accuracy. A significant portion, especially those tackling longitudinal glioma registration, relies on public challenge datasets such as BraTS-Reg, RESECT, BITE, and ReMIND. Table~\ref{tab:datasets_summary} provides comprehensive information on these datasets, which cover a wide variety of imaging protocols (e.g.,~scanner vendor, field-of-strength, resolution, contrast), patient demographics (age, gender, tumor type, grade, location, extent, scope), and annotations (landmarks and segmentations). Such curated datasets provide a common reference for comparing algorithms by supplying standard benchmarks and expert annotations; even so, as presented later in this section, differences in modality pairs, annotation protocols, and preprocessing continue to limit direct per-study comparisons. Several studies also utilize private in-house clinical datasets to address specific neurosurgical scenarios lacking in public challenges. Although these datasets provide highly relevant, real-world clinical data, they limit reproducibility and standardization across studies. 

Moreover, synthetic data are widely employed, either by generating realistic deformations through patient-specific or atlas-based simulations applied to preoperative MRI, or by synthesizing
complementary imaging modalities (e.g.,~ultrasound or CT) from MRI, enabling supervised training when ground truth is unavailable and improving robustness in multimodal settings. However, models
trained on synthetic data may not generalize perfectly to the complexities of real-world clinical data. Several studies have explicitly addressed this issue. \citet{han2022deformable} demonstrated
that their model, trained on simulated data, could successfully be applied to real clinical data, while the pair-specific fine-tuning strategy of \citet{meng2022brain} represents a key approach
for adapting generalized models to patient-specific data. In addition, models often combine multiple datasets during training to increase variability and improve robustness
\citep{canalini2020enhanced,zeineldin2021iregnet,chel2023segmentation,rahmani2024d2bgan,aziz2025efficientmorph}. 

Quantitative evaluation of registration methods typically relies on both geometric and intensity-based metrics. Landmark-based TRE is the most common geometric measure, particularly when corresponding points can be annotated pre- and intraoperatively, and the Dice score is widely used to assess overlap between warped labels. Among similarity terms used to drive registration, NCC is the most common in both unimodal and multimodal tasks, while learned similarity criteria, such as GAN-based losses, are increasingly used in place of handcrafted measures to implicitly capture modality-invariant relationships. To assess the physical plausibility of the deformation, the predominant metric is the percentage of voxels with a non-positive Jacobian determinant \hbox{($\%|\boldsymbol{J}_\phi| \leq 0$)} or the number of folded voxels.

The most widely used classic baselines are the iterative methods NiftyReg \citep{modat2014global}, SyN \citep{avants2008symmetric}, and Elastix \citep{klein2009elastix}. Among learning-based
approaches, VoxelMorph \citep{balakrishnan2019voxelmorph} is the standard CNN baseline, while TransMorph \citep{chen2022transmorph} and LapIRN \citep{mok2020large,mok2021conditional} represent
strong Transformer-based and multiresolution pyramidal baselines, respectively. Table~\ref{tab:registration_results} summarizes registration error and runtime across public benchmarks, grouped by
dataset, task, and method class to ensure comparisons are made only where meaningful. State-of-the-art accuracy is typically $<$$1$--$2$~mm on RESECT, $1$--$3$~mm on BITE
\citep{zeineldin2021iregnet,rahmani2024d2bgan,pirhadi2023robust,canalini2020enhanced}, $1$--$3$~mm on \hbox{BraTS-Reg}
\citep{mok2022unsupervised,zhang2024deep,feng2024stepwise,wodzinski2022unsupervised,wu2024noise}, and $3$--$4$~mm on ReMIND (ReMIND2Reg subset) \citep{aziz2025efficientmorph,hansen2025learn2reg}.
On BraTS-Reg, recent DL methods often match or surpass classic approaches, particularly hybrid models \citep{wodzinski2022unsupervised} and masking-based methods
\citep{mok2022unsupervised,wu2024noise,feng2024stepwise,zhang2024deep}; however, on the more challenging ReMIND2Reg setting, classic methods remain competitive or superior
(see Section~\ref{sec:grand_challenges}). Runtime analysis in Table~\ref{tab:registration_results} also shows a clear split, with standard DL designs running in milliseconds to a few seconds, whereas hybrid
and keypoint-based methods that retain an optimization or detection step stay in the tens-of-seconds to minutes range, comparable to classic pipelines (Section~\ref{sec:disc_takeaways}).

\begin{table*}[t!]
\caption{Reporting quality across the reviewed methods ($n = 46$). ``Robustness'' is reported over the learning-based methods only ($n = 33$). Each bar shows the percentage of publications that address (colored) versus do not address (gray) a given item. Subcategories within each domain are not mutually exclusive; a single publication may contribute to more than one row. Details for each study can be found in Table~\ref{tab:methods_details}.}
\label{tab:reporting_quality}
\fontsize{8}{10}\selectfont
\begin{tabular}{p{0.18\linewidth}p{0.24\linewidth}p{0.51\linewidth}}
\toprule
\textbf{Domain} &
\textbf{Subcategory} &
\textbf{Percentage of Publications} \\
\midrule
\multirow{8}{=}{\textbf{Robustness}}
  & Multi-institutional data        & \minibar{robcol}{17}{33}{52} \\
  & K-fold cross-validation         & \minibar{robcol}{10}{33}{30} \\
  & Data augmentation               & \minibar{robcol}{8}{33}{24} \\
  & Cross-dataset evaluation        & \minibar{robcol}{8}{33}{24} \\
  & Pretraining                     & \minibar{robcol}{5}{33}{15} \\
  & Out-of-distribution evaluation  & \minibar{robcol}{2}{33}{6}  \\
  & Dropout layers                  & \minibar{robcol}{1}{33}{3}  \\
  & Network pruning                 & \minibar{robcol}{1}{33}{3}  \\

\midrule
\multirow{5}{=}{\textbf{Statistical Testing}}
  & Paired t-tests                  & \minibar{statcol}{10}{46}{22} \\
  & Wilcoxon signed-rank tests      & \minibar{statcol}{5}{46}{11} \\
  & Wilcoxon rank-sum tests         & \minibar{statcol}{4}{46}{9}  \\
  & ANOVA tests                     & \minibar{statcol}{2}{46}{4}  \\
  & Tukey's range tests             & \minibar{statcol}{1}{46}{2}  \\

\midrule
\multirow{3}{=}{\textbf{Uncertainty and Bias}}
  & Probabilistic displacement maps & \minibar{unccol}{4}{46}{9}  \\
  & Bias assessment                 & \minibar{unccol}{4}{46}{9}  \\
  & Registration error estimation   & \minibar{unccol}{3}{46}{7}  \\

\midrule
\multirow{4}{=}{\textbf{Reproducibility}}
  & Preprocessing described        & \minibar{repcol}{40}{46}{87} \\
  & Public data used               & \minibar{repcol}{34}{46}{74} \\
  & Hardware reported              & \minibar{repcol}{30}{46}{65} \\
  & Public code available          & \minibar{repcol}{13}{46}{28} \\

\bottomrule
\end{tabular}
\vspace{0.1em}
\\
\footnotesize
ANOVA --- Analysis of Variance.
\end{table*}

The manner in which methods are validated varies considerably across the literature. Most studies perform validation on held-out test sets drawn from the same dataset as the
training data, with cross-validation strategies employed occasionally, especially under limited sample sizes. External validation through cross-dataset validation or OOD clinical data is rare
\citep{baheti2024brain} but is particularly valuable for demonstrating generalizability and domain transfer, which are critical for clinical deployment. In some works, especially those using
simulated data, qualitative inspection by clinical experts or visual overlay comparisons are used to complement quantitative metrics; however, few papers report inter-rater variability or use
standardized expert evaluation protocols. Challenge reports such as those of \hbox{Learn2Reg} \hbox{ReMIND2Reg} \citep{hering2022learn2reg,hansen2025learn2reg}, BraTS-Reg \citep{baheti2024brain}, and
CuRIOUS \citep{xiao2019evaluation} provide valuable statistical benchmarks and facilitate fair comparison, but they remain the exception.

Table~\ref{tab:methods_details} provides, for each reviewed study, a structured assessment of experimental rigor, spanning experimental settings, preprocessing, robustness and generalization strategies, statistical testing, and uncertainty quantification. Table~\ref{tab:reporting_quality} and the following paragraphs summarize the current state of evaluation and reporting across the reviewed studies. Robustness practices are the most commonly adopted, with multi-institutional data being the most frequent ($52$\%), followed by k-fold cross-validation ($30$\%), data augmentation ($24$\%), cross-dataset evaluation ($24$\%), and pretraining ($15$\%), whereas explicit OOD testing remains rare ($6$\%). Uncertainty and bias assessment of methodological choices, results, and predictions is the least addressed category, split among registration error estimation ($7$\%), probabilistic displacement maps ($9$\%), and author discussion of possible biases ($9$\%). Reproducibility is similarly uneven, as preprocessing is described by most studies ($87$\%), with public data ($74$\%) and hardware ($65$\%) also commonly reported, while public code is released by only $28$\% of all methods --- for less than half of DL-based methods (Tables~\ref{tab:dl_methods} and \ref{tab:dl_physics_hybrid_methods}) and almost never for classic (Table~\ref{tab:classic_methods_summary}) or biomechanical pipelines (Table~\ref{tab:physics_methods}).

To get a complete picture, the preprocessing pipelines across the different studies are compiled in Table~\ref{tab:methods_details}. Most DL-based methods rely on a substantial preprocessing chain, typically combining rigid or affine pre-registration, isotropic voxel resampling, intensity normalization, and fixed-size volume cropping, before the network is applied, whereas biomechanical methods instead require anatomical segmentation and mesh generation. These differences in preprocessing make an already difficult head-to-head model comparison even more challenging, which is a point we revisit in Section~\ref{sec:disc_takeaways}. Table~\ref{tab:methods_details} additionally documents the heterogeneity of experimental settings, including dataset sizes and splits, optimizers, and hardware. The consistently small dataset sizes stand out as the main finding, with studies often using only a few ($4$--$30$) cases and rarely exceeding a few hundred. Some methods address this scarcity by synthesizing new data, performing data augmentation, or pretraining on a related task (e.g.,~atlas or inter-patient registration) before fine-tuning. Training configurations are otherwise quite uniform, with most methods relying on the Adam optimizer at a learning rate near $1\times10^{-4}$ with scheduled decay and, for 3D volumetric data, a batch size of $1$ dictated by memory constraints on a single workstation-grade GPU.

\section{Grand challenges for brain image registration} 
\label{sec:grand_challenges}
Recent challenges have played an important role in benchmarking and advancing algorithms for deformable registration in complex neurosurgical and longitudinal imaging settings. Two of the most
relevant initiatives in this context are the BraTS-Reg challenge \citep{baheti2024brain} and the \hbox{ReMIND2Reg} task \citep{dorent2025brain} from the Learn2Reg challenge
\citep{hansen2025learn2reg}. Both focus on pathology-driven deformation modeling, where missing correspondences and large tissue displacements make conventional registration particularly
difficult, differing in the surgical time points and imaging modalities involved.

\subsection{BraTS-Reg challenge (ISBI/MICCAI 2022)}
\label{sec:bratsreg}
The BraTS-Reg (2022) challenge established the first large-scale public benchmark for deformable registration between preoperative and follow-up MRI scans of patients with diffuse gliomas. It aimed to evaluate algorithms under realistic clinical conditions using multiparametric MRI data (T1, ceT1, T2, and T2-FLAIR) from $259$ multi-institutional patients, comprising $160$ training pairs, $50$ testing pairs, and a further $49$ pairs withheld for OOD evaluation. Ground-truth correspondences were provided via expert-annotated landmarks, and submissions were evaluated using Median Euclidean Error (MEE), robustness, and the percentage of voxels with non-positive Jacobian determinant (the last not used for method ranking).

Most of the top-performing methods have already been described in Section~\ref{sec:missing_correspondences}, and Table~\ref{tab:challenges_results}-A reports the official challenge rankings of the leading
submissions. These methods jointly employed pre-alignment, deep neural networks, inverse-consistency analysis, and per-case test-time instance optimization as a post-processing refinement stage to
address missing correspondences caused by tumor resection. In addition to those methods already discussed, another top-ranked submission by \citet{grossbrohmer2022employing} combined MIND-SSC features
with a two-stage optimization strategy, yielding smooth and inverse-consistent deformations.

The BraTS-Reg results highlighted that hybrid registration strategies combining DL with an IO refining step are some of the most effective methods for addressing resection-induced brain
deformation. A
notable observation is that the classic ANTs (SyN) baseline \citep{avants2008symmetric,avants2011reproducible} performed on par with the top learning-based submissions, especially on the OOD
cohort, where the leading methods were statistically indistinguishable from one another ($p > 0.05$). Further details about the challenge and datasets are available in~\citet{baheti2024brain} and on the
official challenge website\footnote{\url{https://www.med.upenn.edu/cbica/brats-reg-challenge/}}.

\begin{table}[!t]
\caption{Quantitative results of selected top-performing methods on the testing sets from the (A) BraTS-Reg (combined ISBI and MICCAI) and (B) ReMIND2Reg challenges. Teams are ordered by the official challenge ranking. Asterisk (*) indicates baseline methods for comparison. ``D'' indicates deep learning-based methods, ``C'' indicates classic methods, and ``H'' indicates a hybrid method that uses both.}
\label{tab:challenges_results}
\resizebox{\columnwidth}{!}{%
\begin{tabular}{lcccc}
\toprule
\multicolumn{5}{c}{\textbf{(A) BraTS-Reg$^a$ \citep{baheti2024brain}}} \\ \midrule
\multirow{2}{*}{\textbf{Team}} & \multicolumn{2}{c}{\textbf{MEE $\downarrow$ (mm)}} & \multicolumn{2}{c}{\textbf{Robustness $\uparrow$}}    \\ \cline{2-5} 
 & \multicolumn{1}{l}{\raisebox{-.5ex}{Median}} & \multicolumn{1}{l}{\raisebox{-.5ex}{Mean}} & \multicolumn{1}{l}{\raisebox{-.5ex}{Median}} & \multicolumn{1}{l}{\raisebox{-.5ex}{Mean}} \\ \midrule
DIRAC \citep{mok2022unsupervised} (D)                            & 1.60     & 1.73    & 0.90     & 0.87       \\
\citet{wodzinski2022unsupervised} (H)                            & 1.66     & 1.79    & 0.90     & 0.86       \\
\citet{grossbrohmer2022employing} (C)                            & 1.71     & 1.79    & 0.90     & 0.87       \\
\citet{canalini2022iterative} (C)                                & 1.59     & 1.95    & 0.90     & 0.87       \\
ANTs* \citep{avants2011reproducible} (C)                         & 1.78     & 2.10    & 0.81     & 0.83       \\
GradICON* \citep{tian2023gradicon} (D)                           & 1.95     & 1.97    & 0.86     & 0.83       \\
SynthMorph* \citep{hoffmann2021synthmorph} (D)                   & 2.15     & 2.36    & 0.80     & 0.82       \\
NICE-Net \citep{meng2022brain} (D)                               & 2.38     & 2.68    & 0.89     & 0.85       \\
iRegNet \citep{zeineldin2022self} (D)                            & 2.76     & 3.17    & 0.80     & 0.78       \\
\citet{abderezaei20223d} (D)                                     & 2.99     & 4.26    & 0.80     & 0.79       \\
\rowcolor{LightGray}
Affine                                                           & 3.59     & 4.14    & 0.61     & 0.61       \\
WSSAMNet \citep{almahfouz2022wssamnet} (D)                       & 4.62     & 5.96    & 0.10     & 0.19      \\ 
\vspace{-0.3em} \\
\toprule
\multicolumn{5}{c}{\textbf{(B) ReMIND2Reg$^a$ \citep{hansen2025learn2reg}}} \\ \midrule
\textbf{Team} & \multicolumn{2}{c}{\textbf{TRE $\downarrow$ (mm)}} & \multicolumn{2}{c}{\textbf{TRE30 $\downarrow$ (mm)}} \\ \midrule
NiftyReg* \citep{modat2014global} (C)           & \multicolumn{2}{c}{2.87} & \multicolumn{2}{c}{4.54} \\
junyi-wang (H)                                  & \multicolumn{2}{c}{4.42} & \multicolumn{2}{c}{6.01} \\
VROC  (C)                                       & \multicolumn{2}{c}{3.63} & \multicolumn{2}{c}{6.24} \\
next\_gen\_nn \citep{wang2024unsupervised} (C)  & \multicolumn{2}{c}{3.71} & \multicolumn{2}{c}{6.49} \\
\rowcolor{LightGray}
Initial error                                   & \multicolumn{2}{c}{4.81} & \multicolumn{2}{c}{7.50} \\
\bottomrule
\end{tabular}
}

\noindent \\ \footnotesize MEE --- Median Euclidean Error; TRE --- Target Registration Error. \\ 
$^a$ Please refer to the official challenges for additional details on rankings, statistical analysis, and methodologies.
\end{table}

\subsection{ReMIND2Reg challenge (Learn2Reg 2024)}
\label{sec:remind2reg}
The ReMIND2Reg (2024) challenge addresses one of the most demanding scenarios in neurosurgical image registration: the multimodal alignment of preoperative MRI and post-resection iUS. Using data
derived from ReMIND \citep{juvekar2024remind} from Brigham and Women's Hospital, it provided $99$ paired scans (3D iUS with ceT1 and/or T2 MRI) for training, four for validation, and $10$ for testing.
Submitted methods needed to manage missing modalities, severe tissue deformation, resection cavities, and inherent US artifacts. Performance was evaluated using manually annotated landmarks via
TRE and TRE30 (the TRE computed on the $30$\% most misaligned landmarks before registration).

Thirteen teams participated in the validation phase, with five reaching the final test stage, reflecting the task's complexity. The best overall performance was achieved by the organizers'
NiftyReg baseline, a classic symmetric block-matching method \citep{modat2014global}, which outperformed all submissions. Among the competing teams, the top-ranked submission used a learning-based
component, combining CycleGAN \hbox{US-to-MRI} synthesis with coarse-to-fine deformable registration, whereas the second team relied on classic variational registration, and the third team on pyramidal convex optimization with MIND-SSC features \citep{wang2024unsupervised,wang2026unsupervised}.
Quantitative results are summarized in Table~\ref{tab:challenges_results}-B. Further details are available in \citet{hansen2025learn2reg} and on the official challenge
website\footnote{\url{https://learn2reg.grand-challenge.org/}}.

Consistent with the findings of BraTS-Reg, the \hbox{ReMIND2Reg} results reaffirmed that traditional optimization methods remain strong baselines for post-resection registration tasks. The challenge also highlighted the limitations of current learning-based methods under severe correspondence loss and modality disparities. Further progress to solve this task will likely depend on hybrid learning frameworks supported by larger annotated datasets to enable deep neural networks to reach the high accuracy required for clinical translation.

\section{Discussion}
\label{sec:discussion}
In this review, we examined recent computational methods for brain deformation compensation in IGN. From an initial pool of $817$ records identified across multiple databases, $46$ studies met the inclusion criteria, which specifically targeted methods addressing deformations directly induced by neurosurgical interventions. This excluded most works performing inter-patient, atlas-based, or non-resection registration, as these do not capture the complex, topology-altering deformations occurring during surgery. Research activity in this area has increased since 2020, with a marked rise following the release of the BraTS-Reg challenge in 2022 \citep{baheti2024brain}. Compared with earlier benchmarks such as the CuRIOUS 2018 challenge \citep{xiao2019evaluation}, BraTS-Reg offered a substantially larger volume of data, albeit limited to MRI, with more thorough benchmarking. Together with the rapid adoption of DL, this produced a clear surge in learning-based registration research, as DL now dominates the literature, representing nearly three-quarters of the reviewed studies. This dominance, however, does not necessarily translate into superior performance, particularly in multimodal registration, where, under standardized challenge evaluation, classic methods often remain competitive with or ahead of learning-based ones.

A unified comparison across all included studies would risk overstating the comparability of the current literature, since methods target different sub-problems, modality pairs, metrics, and annotation protocols. Therefore, we performed a structured synthesis across a predefined set of extraction fields (Table~\ref{tab:methods_details}), complemented by benchmark-specific quantitative summaries when possible (Tables~\ref{tab:registration_results} and \ref{tab:challenges_results}). The takeaways below are organized by registration sub-problem, but they map directly onto the central challenges that recur across the review, including the multimodal domain gap, missing correspondences after resection, and physical plausibility under large deformation. For each, we summarize what the methods do well, where they break down, and the conditions under which they are the preferred choice to explore.

\subsection{Key methodological takeaways}
\label{sec:disc_takeaways}
The reviewed literature indicates that intraoperative brain registration remains an active area of research. Progress varies across its constituent sub-problems, with some approaches demonstrating
greater methodological maturity and clinical validation than others. Direct displacement field regression is usually performed on VoxelMorph-style backbones \citep{balakrishnan2019voxelmorph} and
handles small same-modality deformations but fails when resection cavities break correspondence between image pairs \citep{mok2022unsupervised,canalini2020enhanced,pan2025metamorphic}. Later
methods targeted these failures, building on a small set of shared backbones. For instance, the LapIRN multiresolution pyramid \citep{mok2020large,mok2021conditional} has been widely adopted as a state-of-the-art backbone and built upon by novel methods. In the BraTS-Reg challenge, leading methods took advantage of it and included a forward--backward consistency masking that handles
the missing correspondences left by resection \citep{mok2022unsupervised,wodzinski2022unsupervised,zhang2024deep,feng2024stepwise,tang2024deformable}. Transformer backbones
\citep{chen2022transmorph} have also been adapted for efficient high-resolution registration \citep{aziz2025efficientmorph}. Notably, progress across the field is uneven, with MRI-MRI
registration comparatively well-established, whereas intraoperative multimodal alignment, such as preMRI to post-resection iUS, remains a major open challenge. There, GAN- and VAE-based
cross-modal synthesis \citep{rahmani2024d2bgan,dorent2025unified} is the most active direction, though classic methods remain competitive under standardized evaluation (Section~\ref{sec:grand_challenges}).

\textbf{Direct displacement field regression.} Simply regressing the displacement field is the most straightforward formulation for brain registration using deep learning
(Section~\ref{sec:direct_regression}). Unlike methods that require multiple network passes, such as image synthesis or keypoint matching, it estimates a deformation in a single forward pass over the
moving and fixed images, and this simplicity and reproducibility are why VoxelMorph-derived backbones \citep{balakrishnan2019voxelmorph} recur throughout the literature
\citep{zeineldin2021iregnet,zeineldin2022self,shimamoto2023precise}. The dense voxel-wise output is also where these methods are constrained, as the discontinuities introduced by resection
cavities conflict with the smoothness and topology-preserving assumptions of the deformation model. Accuracy is further limited near these regions, degrading sharply without explicit handling of
missing correspondences (Section~\ref{sec:missing_correspondences}) or diffeomorphic parameterizations. For tasks with small deformations and matching modality pairs, however, these methods perform well
enough to remain the preferred choice.

\textbf{Cross-modality alignment.} Cross-modality registration, despite considerable methodological effort, remains one of the hardest open problems, particularly for large gaps such as
preMRI-to-iUS. Synthesis-driven pipelines try to address this by translating images into a shared appearance space before alignment, using GANs \citep{han2022deformable,rahmani2024d2bgan} or VAEs
\citep{dorent2025unified}, while adversarial losses provide a separate mechanism that replaces handcrafted similarity with a discriminator-learned criterion
\citep{wodzinski2021adversarial,zhu2025synmse} (Section~\ref{sec:gan_synthesis_based}). Their main weaknesses are training instability and a dependence on large paired datasets, which are scarce.
Moreover, synthesis can hallucinate structures where the source modality is uninformative, and the downstream registration inherits these errors. 

In addition to cross-modal synthesis, modality fusion tries to take advantage of complementary information found in different representations of the same tissue, such as in multiparametric MRI.
Methods that implement modality fusion show that the heterogeneity of multiparametric MRI exposes complementary aspects of the pathology. For instance, by combining complementary contrasts
\citep{abderezaei20223d} or anisotropic views \citep{liu2023multi}, methods yielded measurable gains over single-channel baselines, with the largest improvements on cases with complex tumor
appearance changes. However, fusion requires all relevant inputs to be available and aligned, which is not always feasible intraoperatively, where some sequences may be skipped due to time
pressure, cost, or patient safety.

\textbf{Sparse keypoint alignment.} Keypoint-based methods, in contrast to the intensity-based optimization that dominates the field, establish sparse correspondences at distinctive locations
between the moving and fixed images (Section~\ref{sec:landmark_based}). Their main advantage lies in robustness and interpretability, handling large intensity discrepancies between modalities, large
tissue deformations, and topological changes from resection, while allowing the matched keypoints to be directly visualized and assessed \citep{pirhadi2023robust,almahfouz2022wssamnet}. Yet, these
methods are constrained by their reliance on finding good-quality matches for a relatively small set of keypoints, since missed or imperfect matches degrade performance. This limitation is
compounded by their dependence on accurate interpolators to expand sparse correspondences into a dense displacement field for 3D volumes \citep{assis2025deep}.

\textbf{Handling missing correspondences.} Missing correspondences in longitudinal registration are a common and complex problem when dealing with cases of tumor resection or recurrence. Methods
designed for missing correspondences in longitudinal MRI-MRI registration are the most developed contributions in this review. Most adjust the registration objective so that voxels without a
counterpart stop contributing to the loss, whether through hard masks derived from forward--backward consistency \citep{mok2022unsupervised,zhang2024deep}, attention weighting
\citep{tang2024deformable,wu2024noise,feng2024stepwise}, or by reconstructing the pathological image \citep{wu2025crrnet} (Section~\ref{sec:missing_correspondences}). Their main limitation is dataset
specificity, as they are developed for preMRI-to-postoperative MRI (postMRI) registration, and most have not been tested on other longitudinal scenarios.

\textbf{Global context through attention.} Following the broader adoption of attention mechanisms in medical imaging, Transformer-based registration extends the effective receptive field to a
global context. Its main advantage lies in modeling correlated displacements across distant cortical regions, which convolutional architectures struggle with \citep{chen2022transmorph}. Yet, these
models are constrained by the memory cost of self-attention, which forces 3D registration methods to adopt efficient approximations to make it feasible \citep{aziz2025efficientmorph}
(Section~\ref{sec:transformer_based}). Validation and training are further limited by their demand for larger and more diverse training sets than the ones that are currently available, although
fine-tuning pretrained Transformer-based models has proven highly effective and could help overcome this data scarcity \citep{xiao2023mae}.

\textbf{Per-case refinement through hybrid learning.} Hybrid DL+IO frameworks are among the top performers on the BraTS-Reg benchmark (Section~\ref{sec:grand_challenges}). The value of the IO stage is
per-case adaptation, since a network trained on a fixed data distribution cannot match an optimizer that minimizes the residual for the specific image pair at hand (Section~\ref{sec:hybrid_learning}).
The benefits of this refinement are most beneficial near the resection cavity, where the largest deformations occur and correction matters the most
\citep{wodzinski2022unsupervised,waldmannstetter2023primitive,ha2021modality}. However, these methods are constrained by the added cost of the optimization stage, as a multiresolution IO step
substantially increases runtime. Validation in the clinic is further limited because, while this runtime may remain usable, it may not allow for real-time registration.

\textbf{Physics-guided registration.} Despite significant advancements, robust biomechanical models of brain deformation
\citep{frisken2020comparison,luo2020accounting,narasimhan2020accounting,drakopoulos2021adaptive} remain largely research tools with limited clinical adoption. Their main advantage lies in physical
interpretability, explicitly modeling tissue mechanics and BCs to yield anatomically plausible deformations, even with sparse intraoperative data (Section~\ref{sec:bio_constrained}). However, high
computational cost, sensitivity to meshing and segmentation quality, and dependence on uncertain patient-specific material properties still constrain these models. Validation is further limited
by the absence of dense ground-truth deformation data, often reduced to sparse landmark- or image-based comparisons. Recent DL approaches, notably PINNs and graph- or point-based networks
\citep{salehi2022physgnn}, have begun to mitigate these issues by embedding physical laws into the learning process. These models eliminate explicit meshing, achieve real-time performance, and
generalize across geometries and BCs while maintaining biomechanical plausibility (Section~\ref{sec:DL_physics}). However, robust estimation of tissue material properties remains a major bottleneck, as
these parameters vary widely across patients and are rarely measurable \textit{in vivo}. Progress toward integrating multimodal intraoperative data to dynamically infer such properties will be key
to achieving real-time, patient-specific, and physics-consistent modeling for IGN.

\subsection{Validation and uncertainty}
\label{sec:disc_validation}
Validation remains the weakest link in the path to clinical adoption. Quantitative assessment relies on average geometric accuracy, such as TRE, HD, and Dice, which captures how well a method aligns anatomy on a benchmark but says little about where a given registration can be trusted. These measures are made harder to compare due to how heterogeneous the evaluation setups are across studies, as most studies report performance on held-out splits of a single dataset and lean on synthetic deformations or simulated cross-modal images for supervision. External validation on cross-dataset or OOD clinical data, which is the setting that most resembles real intraoperative deployment, remains rare: among the learning-based studies, only $24$\% evaluate across datasets and $6$\% test explicitly OOD (Table~\ref{tab:reporting_quality}). As a result, reported accuracy tends to reflect in-distribution performance under favorable preprocessing rather than the complex settings in the OR.

Registration is still treated as a point estimate, but in neurosurgery, residual misalignment is inevitable under pathology, modality disparity, and artifacts. A single deformation field conveys
no indication of its own reliability, and manual verification is impractical mid-procedure. The emerging work on dense error estimation \citep{salari2023dense,salari2023focalerrornet} and on
models that predict uncertainty alongside the deformation \citep{siegert2024pulpo} is therefore a prerequisite for the safe deployment of novel methods, since a surgeon needs to know not only the
predicted correction but where they can trust it. A clear indication that current validation priorities are misaligned with clinical need is that some form of uncertainty or bias assessment
appears in only a few of the reviewed studies, with explicit registration error estimation or probabilistic displacement maps in roughly $7$\% and $9$\%, respectively (Table~\ref{tab:reporting_quality}).

Many of these limitations reflect the relative immaturity of intraoperative registration, which is still largely in a phase of dataset construction and early model development. Only a handful of
annotated datasets exist, and most are (1) small, with studies often training on only $4$--$30$ cases and rarely exceeding a few hundred (Table~\ref{tab:methods_details}); (2)
single-institution, with only half of the learning-based methods integrating multi-institutional data (Table~\ref{tab:reporting_quality}); and (3) confined to specific modality pairs. As a result, the
available data capture only a fraction of the variability encountered in real surgery. The multi-institutional datasets now available have been valuable in enabling the first meaningful
benchmarking efforts, but they remain too few and too narrow to support the standardized protocols, consistent statistical reporting, and external validation that mature evaluation requires. As
larger and more diverse datasets become available \citep{juvekar2024remind}, studies will increasingly be able to demonstrate the reproducibility and trustworthiness that clinicians and regulatory
bodies require, beyond methodological breakthroughs alone.

\subsection{Clinical integration and future directions}
\label{sec:disc_clinical}
In current clinical practice, brain shift compensation is predominantly achieved through conventional registration techniques integrated into commercial navigation platforms, such as those
developed by Brainlab SE (Munich, Germany) and Medtronic plc (Dublin, Ireland). These typically rely on semi-automated alignment between preoperative images and intraoperative modalities such as
iUS and iMRI to restore navigational accuracy, employing classic algorithms based on MI optimization and B-spline deformation models that are favored for their interpretability, computational
efficiency, and established clinical reliability. A recent example is the Brainlab Ultrasound Snap to MRI, which performs semi-automated rigid \hbox{iUS-to-MRI} fusion within a full acquisition
and registration pipeline of approximately $60$~s \citep{mazzucchi_automatic_2023}. Iterative methods such as Demons \citep{thirion1998image} and SyN
\citep{avants2008symmetric,avants2011reproducible} likewise remain widely used in research as strong, well-validated baselines. Alongside these established systems, learning-based deformation
models have been developed to directly complement and extend intraoperative updates, leveraging navigated tool-tip tracking \citep{juvekar2023mapping}, surgical microscope feeds
\citep{haouchine2021pose,haouchine2023learning}, and ultrasound sweep acquisitions \citep{zeineldin2021iregnet,rahmani2024d2bgan,aziz2025efficientmorph} to estimate tissue deformation without
introducing new hardware into the surgical workflow.

Despite promising research performance, learning-based methods have yet to achieve meaningful clinical translation, and several compounding barriers explain why. A major difference is that classic
methods require no training data and optimize each case individually, leaving them largely indifferent to the broader data distribution, whereas DL-based methods must generalize from a finite
training set and are correspondingly more sensitive to distribution shift and to gaps between their training data and the cases seen at deployment. Under realistic clinical conditions, where
noise, motion artifacts, and scanner variability are unavoidable, this makes classic IO methods comparatively more robust, while the apparent advantage of learning-based approaches may be
overestimated relative to real-world clinical settings because validation is confined to retrospective public datasets that are preprocessed, defaced, and affinely pre-registered
\citep{baheti2024brain}. 

Most DL studies also lack prospective clinical validation and cross-institutional generalization, and few provide evidence of clinical benefit based on patient outcomes. Additionally, deep neural
networks provide limited insight into why a particular deformation was predicted or how they might fail, which is difficult to reconcile with the safety and transparency requirements of clinical
deployment and regulatory frameworks such as FDA clearance pathways. On top of these barriers, surgeons have no intuitive way to validate nonrigid updates mid-procedure. Unlike standard rigid
alignments, which can be easily inspected and adjusted, nonrigid deformations are difficult to verify anatomically. Because of this, building tools that let clinicians confidently validate these
updates is just as critical as the algorithm's underlying accuracy \citep{salari2023dense,salari2023focalerrornet,bierbrier2023toward,siegert2024pulpo}.

Overcoming these barriers will depend on several factors. The most impactful next step would likely be a large, multi-institutional benchmark with per-case landmark annotations in the spirit of
BraTS-Reg, but refocused onto preMRI-to-iUS. This pairing reflects the most common clinical reality, given that far more centers have intraoperative ultrasound than intraoperative MRI
\citep{dixon2022intraoperative}. Resources such as ReMIND already provide substantial MRI-US data \citep{juvekar2024remind}, but the absence of landmarks for every case is the limiting factor.
Since no dense ground-truth deformation exists for this problem, landmarks are the only form of supervision that can anchor training to the anatomy encountered in the OR. Lacking such supervision,
unsupervised methods are bounded by the fidelity of their image-similarity metric, which is hard to define across modalities \citep{xu2020adversarial,zou2022review}.

Methodologically, broader access to iUS would not by itself resolve the cross-modal challenge, as the low signal-to-noise ratio that makes ultrasound hard for algorithms to ``interpret'' would persist.
Cross-modal synthesis methods, such as CycleGANs and VAEs, are among the most promising directions, recasting a multimodal problem as an essentially unimodal one in a shared representation space
before registration. Additionally, pretrained and foundation models remain underexplored in medical image registration beyond the unimodal and multimodal developments of GradICON
\citep{tian2024unigradicon,demir2024multigradicon}, again partly because of limited data. Pretraining on simulated data or simpler tasks before fine-tuning on a specific and more complex
registration task offers a way to mitigate the scarcity of annotated data \citep{hoffmann2021synthmorph,zheng2024residual}, which is consistent with the broad success of pretraining across other
fields.

Finally, integration into the OR will ultimately require prospective, multi-institutional clinical validation, alongside interpretable outputs and genuine workflow efficiency
\citep{geshvadi2025optimizing}. In the OR, efficiency is measured in surgeon focus rather than sheer computational speed. Neurosurgery takes hours, so waiting a few seconds for an update is rarely
an issue. The real bottleneck is the cognitive load required to request, evaluate, and ultimately read and trust the system, as its outputs must be conveyed through clear guidance rather than cluttered
overlays that add to the surgeon's visual burden. Rather than replacing established classic approaches, the most viable near-term path is likely through hybrid systems that pair the reliability
and regulatory familiarity of classic methods with the adaptability of learning-based ones, running alongside conventional navigation to build clinical evidence incrementally before broader
deployment. Achieving this transition will require sustained collaboration between engineers, clinicians, and regulatory bodies to ensure that technical advances translate into meaningful
improvements to patient outcomes.


\section*{CRediT authorship contribution statement}
\textbf{Tiago Assis}: Conceptualization, Data curation, Investigation, Methodology, Project administration, Supervision, Validation, Visualization, Writing - original draft, Writing - review \& editing. \textbf{Colin P. Galvin}: Conceptualization, Investigation, Methodology, Validation, Visualization, Writing - original draft, Writing - review \& editing. \textbf{Joshua P. Castillo}: Conceptualization, Investigation, Methodology, Validation, Visualization, Writing - original draft, Writing - review \& editing. \textbf{Nazim Haouchine}: Conceptualization, Investigation, Visualization, Writing - original draft, Writing - review \& editing. \textbf{Marta Kersten-Oertel}: Conceptualization, Methodology, Supervision, Writing - review \& editing. \textbf{Zeyu Gao}: Writing - review \& editing. \textbf{Mireia Crispin-Ortuzar}: Writing - review \& editing. \textbf{Stephen J. Price}: Writing - review \& editing. \textbf{Thomas Santarius}: Writing - review \& editing. \textbf{Yangming Ou}: Supervision, Writing - review \& editing. \textbf{Sarah Frisken}: Writing - review \& editing. \textbf{Nuno C. Garcia}: Conceptualization, Supervision, Writing - review \& editing. \textbf{Alexandra J. Golby}: Supervision, Writing - review \& editing. \textbf{Reuben Dorent}: Conceptualization, Supervision, Writing - original draft, Writing - review \& editing. \textbf{Ines P. Machado}: Conceptualization, Investigation, Methodology, Project administration, Supervision, Validation, Visualization, Writing - original draft, Writing - review \& editing.

\section*{Declaration of competing interest}
The authors declare that they have no known competing financial interests or personal relationships that could have appeared to influence the work reported in this paper.

\section*{Acknowledgements}
The work of Tiago Assis and Nuno C. Garcia was supported by Fundação para a Ciência e a Tecnologia (FCT) through the LASIGE Research Unit, Ref. UID/408/2025. The work of Nazim Haouchine was supported by the National Institutes of Health (NIH) Grants K25EB035166 and R03EB033910. The work of Yangming Ou was supported by the NIH Grants R33NS126792 and R21NS121735, and the St. Baldrick's Career Development Award. The work of Sarah Frisken was supported by the NIH Grants R01EB034223 and R01EB032387. The work of Alexandra J. Golby was supported by the NIH Grants R01EB032387 and R01NS125781. The work of Reuben Dorent was supported by the Marie Skłodowska-Curie Grant No. 101154248 (Project: SafeREG) and by ANR through the “Investissements d’avenir” program (ANR-10-IAIHU-06, ANR-19-P3IA-0001, PRAIRIE 3IA Institute) and the “France 2030” program (ANR-23-IACL-0008, PRAIRIE-PSAI) project. The work of Mireia Crispin-Ortuzar was supported by the Joseph Mitchell Cancer Research Fund, the Academy of Medical Sciences (G117526), and NIHR (NIHR206092). We acknowledge funding and support from Cancer Research UK and the Cancer Research UK Cambridge Centre (CTRQQR-2021-100012), The Mark Foundation for Cancer Research (RG95043), and GE HealthCare. Additional support was provided by the National Institute of Health Research (NIHR) Cambridge Biomedical Research Centre (NIHR203312). The views expressed are those of the author(s) and not necessarily those of the NIHR or the Department of Health and Social Care.

\appendix
\section{Supplementary Tables}

See Tables~\ref{tab:methods_details} and \ref{tab:datasets_summary}.

\clearpage
\onecolumn

\begin{center}
\fontsize{8}{10}\selectfont
\setlength{\LTcapwidth}{\textwidth}
\begin{longtable}{p{3.4cm}p{3.1cm}p{2.7cm}p{2.7cm}p{2.0cm}p{2.0cm}}
\caption{Training and evaluation settings for reviewed methods by category. ``N/A'' indicates missing or non-applicable information due to varying reporting across studies.} 
\label{tab:methods_details} \vspace{-1em} \\
\toprule
\toprule
\textbf{Method} & \textbf{Experimental Settings} & \textbf{Data Preprocessing} & \textbf{Robustness} & \textbf{Statistical \newline Testing} & \textbf{Uncertainty \newline and Bias} \\
\midrule
\endfirsthead
\multicolumn{6}{c}{\tablename~\thetable{} -- \textit{Continued from previous page.}} \\
\toprule
\textbf{Method} & \textbf{Experimental Settings} & \textbf{Data Preprocessing} & \textbf{Robustness} & \textbf{Statistical \newline Testing} & \textbf{Uncertainty \newline and Bias} \\
\midrule
\endhead
\midrule
\multicolumn{6}{r}{\textit{Continued on next page}} \\
\endfoot
\endlastfoot

\multicolumn{6}{c}{\textbf{CLASSIC INSTANCE OPTIMIZATION}} \\

\textbf{iUS-to-iUS} &  &  &  &  &  \\
\citet{canalini2020enhanced} & 30 patients & Rigid registration & Multi-institutional data & Wilcoxon \newline signed-rank tests & Bias assessment \newline of results \\
\addlinespace[6pt]
\citet{chel2023segmentation} & 43 image pairs & Hyperechoic region \newline extraction & Multi-institutional data & N/A & N/A \\
\textbf{preMRI-to-iUS} &  &  &  &  &  \\
\citet{ghose2021automatic} & 22 patients & Voxel spacing \newline resampling; \newline Sobel filtering; \newline Volume cropping & N/A & N/A & N/A \\
\addlinespace[6pt]
\citet{li2026mrf} & 37 patients; \newline Intel Xeon Gold 6226R $\times$2 & Rigid registration; \newline Isotropic voxel spacing ($0.5$ mm$^3$) & Multi-institutional data; \newline Cross-dataset evaluation & Wilcoxon \newline rank-sum tests & N/A \\
\addlinespace[6pt]
\textbf{preMRI-to-postMRI} &  &  &  &  &  \\
\citet{canalini2022iterative} & 140 patients \newline (hyperparameter search); \newline 20 patients \newline (validation) & N/A & Multi-institutional data & N/A & N/A \\
\addlinespace[6pt]
\textbf{Thermography} &  &  &  &  &  \\
\citet{iorga2023robust} & 10 patients & Temporal \newline downsampling & N/A & Wilcoxon \newline signed-rank tests & N/A \\

\midrule
\multicolumn{6}{c}{\textbf{DEEP LEARNING}} \\

\textbf{preMRI-to-iMRI} &  &  &  &  &  \\
MetaRegNet \newline \citep{joshi2023metaregnet} & 240 train, 5 val, 20 test \newline (image pairs); \newline Adam optimizer; \newline Learning rate $1\times10^{-4}$; \newline Nvidia Titan X 12 GB & Affine registration; \newline Isotropic spacing \newline ($1$ mm$^3$); \newline Min-max normalization & N/A & N/A & N/A \\
\addlinespace[6pt]
\citet{shimamoto2023precise} & 248 patients & Rigid registration; \newline Isotropic voxel spacing \newline ($1$ mm$^3$); \newline Voxel cropping \newline ($192\times192\times192$); \newline Image downsampling \newline ($64\times64\times64$); \newline Max/min pooling; \newline Frequency filtering & 5-fold cross-validation & Wilcoxon \newline signed-rank tests & N/A \\
\addlinespace[6pt]
\textbf{preMRI-to-iCT} &  &  &  &  &  \\
\citet{han2022deformable} & 400 train, 10 val, 10 test \newline (synthetic deformations); \newline Adam optimizer; \newline Learning rate $2\times10^{-4}$; \newline Batch size 1; \newline Nvidia Quadro 6000 24 GB & Rigid registration; \newline Volume cropping \newline ($128\times160\times128$); \newline Isotropic voxel spacing \newline ($1.5$ mm$^3$); \newline Min-max normalization & Cross-dataset and \newline out-of-distribution \newline evaluation & Paired t-tests & Probabilistic \newline predictions \\
\addlinespace[6pt]
\textbf{iUS-to-iUS} &  &  &  &  &  \\
\citet{wodzinski2021adversarial} & 13 train, 1 val, 3 test \newline (patients); \newline Learning rate $2\times10^{-4}$ & Affine registration; \newline Image downsampling \newline ($0.5\times$); \newline Gaussian filtering & 6-fold cross-validation; \newline Data augmentation & N/A & N/A \\
\addlinespace[6pt]
\citet{pirhadi2023robust} & 1260 train, 420 val, 420 test \newline (image pairs); \newline Adam optimizer; \newline Learning rate $5\times10^{-5}$; \newline Batch size 8; \newline Nvidia GTX 1050 Ti 4 GB & Isotropic voxel spacing \newline ($0.14$ mm$^3$); \newline Volume cropping; \newline Min-max normalization & Data augmentation; \newline Pretraining; \newline 5-fold cross-validation; \newline Cross-dataset \newline evaluation & Wilcoxon \newline rank-sum tests & N/A \\
\addlinespace[6pt]
\textbf{preMRI-to-iUS} &  &  &  &  &  \\
EfficientMorph \newline \citep{aziz2025efficientmorph} & 155 train, 10 val, 40 test \newline (image pairs); \newline Adam optimizer; \newline Learning rate $5\times10^{-4}$; \newline Batch size 1; \newline Nvidia A100 40 GB & Rigid registration; \newline Volume cropping \newline ($256\times256\times256$); \newline Isotropic voxel spacing \newline ($0.5$ mm$^3$) & N/A & N/A & N/A \\
\addlinespace[6pt]
D2BGAN \newline \citep{rahmani2024d2bgan} & 264 train, 110 val \newline (image pairs); \newline Learning rate $1\times10^{-5}$ \newline (exponentially decaying); \newline Batch size 24; \newline Nvidia RTX 3060Ti 8 GB & N/A & Cross-dataset \newline evaluation; \newline Multi-institutional data & Paired t-tests & Probabilistic \newline predictions \\
\addlinespace[6pt]
iRegNet \newline \citep{zeineldin2021iregnet} & 23 train, 8 val \newline (patients); \newline Adam optimizer; \newline Learning rate $1\times10^{-4}$; \newline Batch size 2; \newline Nvidia RTX 2080Ti 11 GB & Affine registration; \newline Volume cropping \newline ($128\times128\times128$); \newline Isotropic voxel spacing \newline ($1$ mm$^3$); \newline Intensity normalization & Cross-dataset \newline evaluation; \newline Multi-institutional data & N/A & N/A \\
\addlinespace[6pt]
\citet{ha2021modality} & 19 patients; \newline Adam optimizer; \newline Learning rate $2\times10^{-3}$; \newline Nvidia Quadro 6000 24 GB & Isotropic voxel spacing ($0.5$ mm$^3$); \newline Volume cropping \newline ($64\times64\times64$) & 5-fold cross-validation; \newline Network pruning & Paired t-tests; \newline Wilcoxon \newline rank-sum tests & Probabilistic \newline predictions \\
\addlinespace[6pt]
SynMSE \newline \citep{zhu2025synmse} & 14 train, 8 test \newline (patients); \newline Learning rate $1\times10^{-4}$; \newline Nvidia Quadro 8000 48 GB \newline Nvidia RTX 4090 24 GB & Affine registration; \newline Image downsampling \newline ($128\times128\times144$); \newline Isotropic voxel spacing \newline ($1$ mm$^3$); \newline Intensity normalization & Cross-dataset \newline evaluation; \newline Data augmentation & Paired t-tests & N/A \\
\addlinespace[6pt]
\textbf{preMRI-to-postMRI} &  &  &  &  &  \\
DIRAC \newline \citep{mok2022unsupervised} & 122 train, 10 val, 28 test \newline (patients); \newline Adam optimizer; \newline Learning rate $1\times10^{-4}$; \newline Nvidia Titan RTX 24 GB & Image downsampling \newline ($160\times160\times80$); \newline Voxel size resampling \newline ($1.5\times1.5\times1.94$ mm$^3$) & 5-fold cross-validation; \newline Multi-institutional data & N/A & N/A \\
\addlinespace[6pt]
CRRNet \newline \citep{wu2025crrnet} & 140 (5-fold CV) + \newline 425 train, 110 test \newline (patients); \newline Adam optimizer; \newline Learning rate $1\times10^{-4}$; \newline Batch size 1; \newline Nvidia RTX 4090 & Skull stripping; \newline Rigid registration; \newline Image downsampling \newline ($160\times160\times80$); \newline Voxel size resampling \newline ($1.5\times1.5\times1.94$ mm$^3$) & Multi-institutional data; \newline 5-fold cross-validation; \newline Cross-dataset and \newline out-of-distribution \newline evaluation & Paired t-tests & N/A \\
\addlinespace[6pt]
MetaLapIRN \newline \citep{pan2025metamorphic} & 122 train, 10 val, 28 test \newline (patients); \newline AdamW optimizer; \newline Learning rate $1\times10^{-4}$; \newline Batch size 1; \newline Nvidia RTX 3090 24 GB & Intensity normalization; \newline Image downsampling ($160\times160\times80$); \newline Voxel size resampling \newline ($1$ mm$^3$) & Multi-institutional data; \newline 5-fold cross-validation; \newline Pretraining & N/A & N/A \\
\addlinespace[6pt]
PULPo \newline \citep{siegert2024pulpo} & 120 train, 20 val, 20 test \newline (patients) & Image downsampling \newline ($144\times192\times160$) & Multi-institutional data & N/A & Probabilistic \newline predictions; \newline Uncertainty \newline quantification \\
\addlinespace[6pt]
iRegNet \newline \citep{zeineldin2022self} & 112 train, 28 val, 20 test \newline (patients); \newline Adam optimizer; \newline Learning rate $1\times10^{-4}$; \newline Batch size 2; \newline Nvidia RTX 3060 12 GB & Affine registration; \newline Volume cropping \newline ($160\times192\times160$); \newline Z-score normalization & Multi-institutional data & N/A & N/A \\
\addlinespace[6pt]
WSSAMNet \newline \citep{almahfouz2022wssamnet} & 140 train, 20 val \newline (patients); \newline Adam optimizer; \newline Learning rate $1\times10^{-4}$; \newline Unspecified 12 GB GPUs & N/A & Multi-institutional data & N/A & N/A \\
\addlinespace[6pt]
NICE-Net \newline \citep{meng2022brain} & 140 train, 20 val \newline (patients); \newline Adam optimizer; \newline Learning rate $1\times10^{-4}$ \newline (pretraining); \newline Learning rate $1\times10^{-5}$ \newline (fine-tuning); \newline Batch size 1; \newline Nvidia Titan V 12 GB & Volume cropping \newline ($144\times192\times160$); \newline Min-max normalization & Pretraining; \newline Multi-institutional data & N/A & N/A \\
\addlinespace[6pt]
MSF-AR Net \newline \citep{liu2023multi} & 39 train, 9 val \newline (patients); \newline Adam optimizer; \newline Learning rate $3\times10^{-4}$; \newline Batch size 1; \newline Nvidia GTX 1080Ti 11 GB & N/A & N/A & N/A & N/A \\
\addlinespace[6pt]
\citet{waldmannstetter2023primitive} & 240 train, 30 val, 30 test \newline (patients); \newline Learning rate $1\times10^{-4}$ \newline (network regression); \newline Learning rate $1\times10^{-3}$ \newline (instance optimization) & Rigid registration; \newline normalization & Multi-institutional data & Paired t-tests & N/A \\
\addlinespace[6pt]
\citet{feng2024stepwise} & 112 train, 14 val, 14 test \newline (patients); \newline Adam optimizer; \newline Learning rate $1\times10^{-4}$; \newline Batch size 1; \newline Nvidia RTX 4080 16 GB & Image downsampling \newline ($160\times160\times80$) & 5-fold cross-validation; \newline Multi-institutional data & N/A & N/A \\
\addlinespace[6pt]
\citet{zhang2024deep} & 122 train, 20 val, 28 test \newline (patients); \newline Adam optimizer; \newline Learning rate $1\times10^{-4}$; \newline Batch size 1; \newline Nvidia RTX 3080Ti 12 GB & Image downsampling \newline ($160\times160\times80$); \newline Voxel size resampling \newline ($1.5\times1.5\times1.94$ mm$^3$) & Multi-institutional data & Paired t-tests & N/A \\
\addlinespace[6pt]
\citet{tang2024deformable} & 100 train, 12 val, 28 test \newline (patients); \newline Adam optimizer; \newline Learning rate $1\times10^{-4}$; \newline Nvidia RTX A6000 48 GB & N/A & Multi-institutional data & N/A & N/A \\
\addlinespace[6pt]
\citet{wu2024noise} & 286 train, 109 val \newline (patients); \newline Adam optimizer; \newline Learning rate $1\times10^{-4}$; \newline Batch size 1; \newline Nvidia RTX 4090 24 GB & Image downsampling \newline ($160\times160\times80$) & Cross-dataset \newline evaluation; \newline Multi-institutional data & N/A & N/A \\
\addlinespace[6pt]
\citet{abderezaei20223d} & 140 train, 20 val \newline (patients); \newline Adam optimizer; \newline Learning rate $1\times10^{-4}$ \newline (decaying); \newline Batch size 1; \newline Nvidia A40 48 GB & Affine registration; \newline Histogram matching; \newline Z-score normalization & Multi-institutional data & N/A & N/A \\
\addlinespace[6pt]
\citet{wodzinski2022unsupervised} & 140 train, 20 val \newline (patients); \newline Learning rate $3\times10^{-3}$ \newline (exponentially decaying); \newline Nvidia RTX 3090Ti 24 GB & Affine registration; \newline Min-max normalization & Multi-institutional data & N/A & N/A \\
\addlinespace[6pt]
\citet{duan2023towards} & 122 train, 10 val, 28 test \newline (patients) & N/A & 5-fold cross-validation; \newline Cross-dataset \newline evaluation; \newline Multi-institutional data & Paired t-tests & N/A \\
\addlinespace[6pt]
\textbf{Surgical Microscopy} & & & & & \\
\citet{haouchine2023learning} & 6 patients; \newline 1500 synthetic images \newline (per case); \newline Adam optimizer; \newline Learning rate $1\times10^{-3}$ \newline (exponentially decaying); \newline Batch size 8; \newline Nvidia GTX 1070 8 GB & Cortical vessel \newline segmentation; \newline Cortical vessel \newline mesh triangulation & Pretraining; \newline Leave-one-out \newline cross-validation; \newline Data augmentation & N/A & N/A \\

\pagebreak

\multicolumn{6}{c}{\textbf{BIOMECHANICAL MODELING}} \\

\textbf{FEM} & & & & & \\
\citet{li2026fully} & 15 patients; & Brain segmentation; \newline Tetrahedral meshing; \newline Rigid registration & N/A & Wilcoxon \newline signed-rank tests & N/A \\
\addlinespace[6pt]
\citet{frisken2020comparison} & 15 patients; \newline 24 image pairs & Rigid registration; \newline Brain geometry \newline segmentation; \newline Tetrahedral mesh \newline generation & N/A & Paired t-tests & N/A \\
\addlinespace[6pt]
\citet{luo2020accounting} & 6 patients; \newline 2835 simulations; \newline Intel Core i7 @ 3.60 GHz, \newline 8 GB RAM & Brain geometry \newline segmentation; \newline Tetrahedral mesh \newline generation; \newline Rigid registration & N/A & N/A & Bias assessment \newline of operator errors  \\
\addlinespace[6pt]
\citet{narasimhan2020accounting} & 6 patients; \newline 756 simulations (gravity) + \newline 84 simulations (debulking) & Brain geometry \newline segmentation; \newline Tetrahedral mesh \newline generation; \newline Rigid registration & Multi-institutional data & Wilcoxon \newline signed-rank tests; \newline Wilcoxon \newline rank-sum tests & Bias assessment \newline of case selection  \\
\addlinespace[6pt]
\citet{lesage2021viscoelastic} & 4 patients; \newline >400 simulations; \newline Intel Xeon E5-2680 v3, \newline 48/96 parallel CPU cores, \newline 192 GB RAM per node \newline (HPC cluster) & Skull-stripping; \newline Brain geometry \newline segmentation; \newline Gaussian smoothing; \newline Tetrahedral mesh \newline generation & N/A & ANOVA tests; \newline  Tukey's range \newline tests & N/A  \\
\addlinespace[6pt]
\citet{drakopoulos2021adaptive} & 30 patients; \newline Intel Xeon X5690, \newline 96 GB RAM & Skull-stripping; \newline Rigid registration; \newline Brain geometry \newline segmentation; \newline Tetrahedral mesh \newline generation & Multi-institutional data & N/A & N/A \\

\addlinespace[6pt]
\textbf{Meshless} & & & & & \\
\citet{yu2022automatic} & 4 patients; \newline Intel Core i7 @ 2.80 GHz, \newline 16 GB RAM & Skull-stripping; \newline Brain geometry \newline segmentation; \newline Tetrahedral integration \newline grid generation; \newline Fuzzy material property \newline assignment & N/A & N/A & N/A \\

\pagebreak

\multicolumn{6}{c}{\textbf{DEEP LEARNING + BIOMECHANICS}} \\

\textbf{preMRI-to-iMRI} & & & & & \\
PhysGNN \newline \citep{salehi2022physgnn} & 3465 train, 990 val, 495 test \newline (simulations); \newline AdamW optimizer; \newline Learning rate $5\times10^{-3}$ \newline (reduced on plateau); \newline Nvidia Tesla P100 16 GB & Tetrahedral mesh \newline generation & Dropout layers & N/A & N/A \\
\addlinespace[6pt]
\textbf{Surgical Microscopy} & & & & & \\
\citet{haouchine2021predicted} & 4 patients & Brain geometry \newline segmentation; \newline Hexahedral mesh \newline generation & N/A & N/A & N/A \\
\addlinespace[6pt]
\citet{haouchine2021pose} & \textit{Feature extractor}: \newline 1392 train, 464 val \newline (synthetic images); \newline \textit{Pose estimator}: \newline 6300 train, 2100 val \newline (synthetic poses); \newline \textit{Both}: \newline Adam optimizer; \newline Learning rate $1\times10^{-3}$; \newline Batch size 8; \newline Nvidia GTX 1070 8 GB & Brain geometry \newline segmentation; \newline Tetrahedral mesh \newline generation; \newline Vessel centerline \newline extraction & Data augmentation & N/A & Bias assessment \newline of results  \\

\midrule
\multicolumn{6}{c}{\textbf{REGISTRATION UNCERTAINTY ESTIMATION}} \\
\addlinespace[6pt]
\citet{bierbrier2023toward} & 10 train, 2 val, 2 test \newline (patients); \newline Adam optimizer; \newline Learning rate $1\times10^{-4}$; \newline Nvidia RTX 3090 24 GB & ``N3'' intensity \newline normalization; \newline Histogram matching; \newline Rigid/affine registration; \newline Isotropic voxel spacing \newline (1 mm$^3$) & Data augmentation & Three-way ANOVA tests & Registration \newline error estimation \\
\addlinespace[6pt]
\cite{salari2023dense} & 14 train, 4 val, 4 test \newline (patients); \newline AdamW optimizer; \newline Learning rate $1\times10^{-4}$; \newline Batch size 8 & Volume cropping; \newline Isotropic voxel spacing \newline (0.5 mm$^3$) & Pretraining; \newline Data augmentation & N/A & Registration \newline error estimation \\
\addlinespace[6pt]
\citet{salari2023focalerrornet} & 14 train, 4 val, 4 test \newline (patients); \newline Adam optimizer; \newline Learning rate $5\times10^{-5}$; \newline Batch size 64 & Volume cropping; \newline Isotropic voxel spacing \newline (0.5 mm$^3$) & Data augmentation & Paired t-tests & Registration \newline error estimation \\

\bottomrule
\bottomrule
\vspace{0.5pt}
\footnotesize 
\end{longtable}
\end{center}
\twocolumn

\begin{table*}[t!]
\caption{Comprehensive comparison of key publicly available neurosurgery datasets for benchmarking medical image registration algorithms. ``N/A'' indicates that a given aspect was not reported or not applicable.}
\label{tab:datasets_summary}
\fontsize{8}{10}\selectfont
\begin{threeparttable}
\begin{tabularx}{\textwidth}{p{2.6cm}p{3.5cm}p{3.4cm}p{3.5cm}X}
\toprule
\textbf{Dataset} &
  \textbf{RESECT} \newline \citep{xiao2017retrospective} &
  \textbf{BITE} \newline \citep{mercier2012online} &
  \textbf{ReMIND} \newline \citep{juvekar2024remind} &
  \textbf{BraTS-Reg} \newline \citep{baheti2024brain} \\
\midrule
\textbf{Data Sources} &
  St. Olavs University Hospital, Trondheim, Norway &
  Montreal Neurological \newline Institute, Montreal, Canada &
  Brigham and Women's \newline Hospital, Boston, USA &
  Multi-institutional data: \newline TCIA collections \citep{scarpace2016cancer, puchalski2018anatomic, wang2021proteogenomic, liu2024multi} and BraTS private affiliates \\
\midrule
\textbf{Clinical Data} & & & & \\
\addlinespace[6pt]
  No. of Patients (M:F) &
  $23$\tnote{a} &
  $14$ ($9$:$5$) &
  $114$ ($61$:$53$) &
  $259$\tnote{a} \\
  Age (years) &
  $>$$18$\tnote{a} &
  $52.00 \pm 17.05$ ($23$--$76$) &
  $46.83 \pm 14.76$ ($20$--$76$) &
  $>$$18$\tnote{a} \\
  Tumor Volume\tnote{b} (cm$^3$) &
  $40.0 \pm 44.09$ ($1.4$--$165.9$) &
  $34.65 \pm 23.62$ ($0.2$--$79.2$) &
  $24.08 \pm 23.31$ ($0.04$--$138.48$) &
  N/A \\
  Tumor Diagnosis &
  Astrocytoma ($10$) \newline Oligodendroglioma ($10$) \newline Oligoastrocytoma ($2$) \newline Other ($1$) &
  Astrocytoma ($3$) \newline Oligodendroglioma ($3$) \newline Glioblastoma ($8$) &
  Astrocytoma ($33$) \newline Oligodendroglioma ($24$) \newline Glioblastoma ($31$) \newline Metastases ($11$) \newline Other ($15$) &
  N/A \\
  Glioma Type &
  LGG ($23$) &
  LGG ($4$), HGG ($10$) &
  LGG ($34$), HGG ($60$), N/A ($20$) &
  HGG ($259$) \\
\midrule
\textbf{Image Acquisition} & & & & \\
\addlinespace[6pt]
  MRI Scanner &
  $3$ T Magnetom Skyra  (20) \newline $1.5$ T Magnetom Avanto (3) \newline (Siemens) &
  $1.5$ T Signa EXCITE (13) \newline (GE Healthcare) \newline $3$ T Magnetom Trio (1) \newline (Siemens) &
  Preop: Various $1.5$ T/$3$ T/$7$ T scanners \newline Intraop: $3$ T Magnetom Verio \newline (Siemens) &
  N/A \\
  MRI Sequences &
  ceT1, T2-FLAIR &
  ceT1, T2 &
  \tnote{c} T1, ceT1, T2, T2-FLAIR &
  T1, ceT1, T2, T2-FLAIR \\
  MRI Voxel Size &
  $1.0 \times 1.0 \times 1.0$ mm &
  $1.0 \times 1.0 \times 1.0$ mm &
  Variable &
  $1.0 \times 1.0 \times 1.0$ mm \\
  Neuronavigation Unit &
  Sonowand Invite \newline (Sonowand AS) &
  IBIS NeuroNav \newline \citep{mercier2011new} &
  Cranial Navigation \newline (Brainlab) &
  N/A \\
  Tracking System &
  Polaris Spectra \newline (Northern Digital Inc.) &
  Polaris \newline (Northern Digital Inc.) &
  Polaris Spectra for Brainlab \newline (Northern Digital Inc.) &
  N/A \\
  iUS Processing Unit &
  Built-in &
  HDI 5000 \newline (Philips ATL) &
  bk5000 \newline (BK Medical) &
  N/A \\
  iUS Probe &
  12FLA-L (linear) \newline (Sonowand AS) &
  P7-4 (phased) \newline (Philips ATL) &
  N13C5 (curved) \newline  (BK Medical) &
  N/A \\
  iUS Frequency Range &
  $6$--$12$ MHz &
  $4$--$7$ MHz &
  $5$--$13$ MHz &
  N/A \\
  iUS Reconstr. Res. &
  $0.14 \times 0.14 \times 0.14$ mm to \newline $0.24 \times 0.24 \times 0.24$ mm &
  $0.3 \times 0.3 \times 0.3$ mm &
  $0.1 \times 0.1 \times 0.5$ mm &
  N/A \\
\midrule
\textbf{Annotations} & & & & \\
\addlinespace[6pt]
  Landmarks \newline (pairs per patient) &
  $16$--$34$ \newline (pre- v. postresection iUS); \newline $15$--$16$ \newline (preMRI v. preresection iUS); \newline $9$--$16$ \newline (preMRI v. postresection iUS); &
  $19$--$40$ \newline (preMRI v. preresection iUS) &
  ReMIND2Reg subset only \newline (private) &
  $6$--$50$ \newline (pre- v. postMRI) \\
  Segmentations &
  Tumor, cerebral falx, sulci, \newline resection cavity (manual) \newline \citep{behboodi2024open} &
  Tumor (manual) &
  \tnote{c} Whole tumor, residual tumor, prior resection cavity \newline (manual); \newline \tnote{c} Cerebrum, ventricles \newline (automatic) &
  N/A \\
\midrule
\textbf{State-of-the-Art \newline Accuracy} \newline (TRE) & $<1$--$2$ mm \newline \citep{zeineldin2021iregnet, rahmani2024d2bgan, pirhadi2023robust, canalini2020enhanced} & $1$--$3$ mm \newline \citep{zeineldin2021iregnet, rahmani2024d2bgan, pirhadi2023robust, canalini2020enhanced} & $3$--$4$ mm (ReMIND2Reg) \newline \citep{aziz2025efficientmorph, hansen2025learn2reg} & $1$--$3$ mm \newline \citep{mok2022unsupervised, zhang2024deep, feng2024stepwise, wodzinski2022unsupervised, wu2024noise} \\
\bottomrule
\end{tabularx}
\noindent \\ TCIA --- The Cancer Imaging Archive; LGG --- Low-Grade Glioma; HGG --- High-Grade Glioma; Reconstr. Res. --- Reconstruction Resolution.
\begin{tablenotes}
    \footnotesize
    \item[a] This dataset does not provide data regarding patient age or gender.
    \item[b] Tumor volume is quantified based on manually drawn boundaries in the preoperative MRI.
    \item[c] Data not available for every patient. \\
\end{tablenotes}
\end{threeparttable}
\end{table*}

\clearpage

\section*{Data availability}
No data was used for the research described in the article.

\bibliographystyle{elsarticle-harv} 
\bibliography{refs}






\end{document}